\documentclass[english]{article}
\usepackage[T1]{fontenc} \usepackage[latin9]{inputenc} 

\PassOptionsToPackage{normalem}{ulem} 
\usepackage{amsmath,amssymb,babel,cite,color,dsfont,empheq,framed,geometry,graphicx,mathrsfs,needspace,setspace,tabularx,tikz,ulem,url}

\usepackage[capbesideposition=inside,
facing=yes,capbesidesep=quad]{floatrow}

\makeatletter

\let\@fnsymbol\@arabic

\makeatother

\newcommand{\mh}{\mu_{\mathrm{micro}}}
\newcommand{\lh}{\lambda_{\mathrm{micro}}}
\newcommand{\me}{\mu_{e}}
\newcommand{\mc}{\mu_{c}}
\newcommand{\lle}{\lambda_{e}}

\newcommand{\mLc}{\me L_{c}^{2}}

\newcommand{\R}{\mathbb{R}}
\newcommand{\nablau}{\,\nabla u\,}
\newcommand{\p}{{P}}
\newcommand{\nablap}{\nabla \p}
\newcommand{\Curl}{\,\mathrm{Curl}}
\newcommand{\dev}{\, \mathrm{dev}}
\newcommand{\Div}{\mathrm{Div}}
\newcommand{\tr}{\, \mathrm{tr}}
\newcommand{\sym}{\, \mathrm{sym}\,}

\renewcommand{\skew}{\, \mathrm{skew}\,}

\newcommand{\x}{\cdot}

\newcommand{\sig}{\widetilde{\sigma}}
\newcommand{\Sig}{\sigma}
\newcommand{\n}{ n}

\renewcommand{\skew}{\, \mathrm{skew}}

\newcommand{\devsym}{\dev\sym}

\newcommand{\id}{\,\mathds{1}}

\definecolor{Green}{rgb}{0,0.52,0}

\title{\vspace{-1.0cm}First evidence of non-locality in real band-gap metamaterials: determining parameters in the relaxed micromorphic model}

\author{
	Angela Madeo\footnote{Angela Madeo, corresponding author, angela.madeo@insa-lyon.fr, LGCIE, INSA-Lyon, Université de Lyon, 20 avenue	Albert Einstein, 69621, Villeurbanne cedex, France}  \, and 
	Gabriele Barbagallo\footnote{Gabriele Barbagallo, gabriele.barbagallo@insa-lyon.fr, LaMCoS-CNRS \& LGCIE, INSA-Lyon, Universitité de Lyon, 20 avenue Albert Einstein, 69621, Villeurbanne cedex, France} \, and 
Marco Valerio 	d'Agostino\footnote{Marco Valerio d'Agostino, marco-valerio.dagostino@insa-lyon.fr, LGCIE, INSA-Lyon, Université de Lyon, 20 avenue Albert Einstein, 69621, Villeurbanne cedex, France} \\\, and   Luca Placidi\footnote{Luca Placidi, luca.placidi@uninettunouniversity.net, Università Telematica Internazionale UNINETTUNO, Corso Vittorio Emanuele II, 39, 00186 Roma, Italia}
 \, and 	Patrizio Neff\,\footnote{Patrizio Neff, patrizio.neff@uni-due.de, Head of Chair for 	Nonlinear Analysis and Modelling, Fakultät für Mathematik, Universität Duisburg-Essen,  Mathematik-Carrée, Thea-Leymann-Straße 9, 45127 Essen}}

\begin{document}
	\maketitle 
	\addtocounter{footnote}{5} 
	\vspace{0.5cm}	
	\begin{abstract}
In this paper we propose the first estimate of some elastic parameters of the relaxed micromorphic model on the basis of real experiments of transmission of longitudinal plane waves across an interface separating a classical Cauchy material (steel plate) and a phononic crystal (steel plate with fluid-filled holes). A procedure is set up in order to identify the parameters of our model by superimposing the experimentally-based profile of the reflection coefficient (plotted as function of the frequency of the traveling waves) with the analogous profile obtained via simulations based upon the relaxed micromorphic model. We end up with the determination of 5 out of 6 constitutive parameters which are featured by the relaxed micromorphic model in the isotropic case, plus the determination of the micro-inertia parameter. The sixth elastic parameter, namely the Cosserat couple modulus $\mu_{c}$, still remains undetermined, since experimental data concerning the transmission properties of the considered interface for transverse incident waves are not yet available. A fundamental result of the present paper is the estimate of the non-locality intrinsically associated to the underlying microstructure of the metamaterial. As a matter of fact, we appraise that the characteristic length $L_{c}$ measuring the non-locality of the considered phononic crystal is of the order of $1/3$ of the diameter of the considered fluid-filled holes.	
		\end{abstract}
	
	\vspace{1.2cm}
	
	\hspace{-0.55cm}\textbf{Keywords}: complete band-gaps, non-local effects, relaxed micromorphic model, generalized continuum models, multi-scale modeling
	
	\vspace{1.2cm}
	
	\hspace{-0.55cm}\textbf{AMS 2010 subject classification}:  74A10 (stress), 74A30 (nonsimple materials), 74A60 (micromechanical theories),  74E15 (crystalline structure), 74M25 (micromechanics), 74Q15 (effective constitutive equations)
	
	\vspace{1.2cm}
	
	\hspace{-0.55cm}\textbf{to appear in Proceedings of the Royal Society A}
	\newpage
	
	\tableofcontents
	
	\newpage
	
	\addtocontents{toc}{\vspace{-0.2 cm}}
	
\section{Introduction}

Mechanical band-gap metamaterials are suitably engineered microstructured
materials which are able to inhibit elastic wave propagation in specific
frequency ranges due to the presence of their underlying microstructure.
These frequency intervals at which wave inhibition takes place are
known as frequency band-gaps and their intrinsic characteristics (characteristic
values of the gap frequency, extension of the band-gap, etc.) strongly
depend on the metamaterial microstructure. Such unorthodox dynamical
behavior can be related to two main physical phenomena occurring at
the micro-level: 
\begin{itemize}
	\item local resonance phenomena (Mie resonance): the micro-structural components,
	excited at particular frequencies, start oscillating independently
	of the matrix thus capturing the energy of the propagating wave which
	remains confined at the level of the microstructure. Macroscopic wave
	propagation thus results to be inhibited,
	\item micro-diffusion phenomena (Bragg scattering): when the propagating
	wave has wavelengths which are small enough to start interacting with
	the microstructure of the material, reflection and transmission phenomena
	occur at the micro-level that globally result in an inhibited macroscopic
	wave propagation. 
\end{itemize}
Such resonance and micro-diffusion mechanisms (usually a mix of the
two) are at the basis of both electromagnetic and elastic band-gaps
(see e.g. \cite{armenise2010phononic,lucklum2012two} and they are manifestly related to the
particular microstructural topologies of the considered metamaterials.
In fact, it is well known (see e.g. \cite{armenise2010phononic,liu2000locally,spadoni2009phononic,steurer2007photonic,man2013photonic}) that the
characteristics of the microstructures strongly influence the macroscopic
band-gap behavior.

\medskip{}

In recent works \cite{madeo2015wave,madeo2014band} we proposed a new generalized continuum
model, which we called \textit{relaxed micromorphic} which is able
to account for the onset of microstructure-related frequency band-gaps
\cite{madeo2015wave,madeo2014band} while remaining in the macroscopic framework of continuum
mechanics. Well posedness results have already been proved for this
model \cite{ghiba2014relaxed,neff2014unifying}. On the basis of the results obtained in our previous
works, we can claim that the relaxed micromorphic model is the only
macroscopic continuum model known to date which is simultaneously
able to account for 
\begin{itemize}
	\item prediction of complete band-gaps in mechanical metamaterials 
	\item non-local effects (via the introduction of higher order derivatives
	of the micro distortion tensor in the strain energy density).
\end{itemize}
In \cite{madeo2016reflection} we presented a comprehensive study of jump conditions
that can be imposed at surfaces of discontinuity of the material properties
in relaxed micromorphic media, so establishing a strong basis for
the systematic study of reflection and transmission phenomena in real
band-gap metamaterials. In this paper, we will show that the particular
constraint introduced in \cite{madeo2016reflection} that we called ``\textit{macro
	internal clamp with free microstructure}'' is indeed able to reproduce
real situations in which a Cauchy material (e.g. steel) is connected
to a phononic crystal (e.g. a steel plate with fluid-filled holes).
To give a more technologically oriented taste to our investigations,
we consider here the experimental investigations presented in \cite{lucklum2012two}
in which transmission of longitudinal plane waves at an interface
between a steel plate and a phononic crystal is studied. The two main
aims of the present paper can be identified as follows:
\begin{itemize}
	\item determine, by inverse approach, the maximum possible number of
	constitutive elastic parameters featured by the relaxed micromorphic
	model in the isotropic case by direct comparison with the results proposed in \cite{lucklum2012two} and based on real experiments on specific phononic crystals,
	\item give a first evidence of non-local effects in band-gap metamaterials,
	by quantifying them through the determination of the characteristic
	length $L_{c}$ for the phononic crystal experimentally studied in
\cite{lucklum2012two}.
\end{itemize}

\medskip{}
The relaxed micromorphic model is, by its own nature, a ``macroscopic'' model, in the sense that all the constitutive parameters introduced take into account the presence of the micro-structure in an ``averaged'' sense. Nevertheless, it would be interesting to validate the estimate of the parameters of the relaxed micromorphic model performed in this paper against more ``homogenization-oriented'' methods of the type presented in \cite{avila2008multiscale,sridhar2016homogenization}.

The present paper is organized according to the following structure:
\begin{itemize}
	\item In section 2 we briefly recall the bulk governing equations and the
	associated boundary conditions that have to be used for modeling the
	mechanical behavior of Cauchy continua and of relaxed micromorphic
	media \cite{madeo2015wave,madeo2014band,ghiba2014relaxed,neff2014unifying}. The hypothesis of plane wave is introduced
	and a short discussion concerning the behavior of the dispersion relations
	obtained by means of our relaxed micromorphic model is performed.
	No redundant details about the explicit derivation of such dispersion
	relations are given, for this the reader is referred to \cite{madeo2016reflection}.
	It is nonetheless explicitly pointed out that the relaxed micromorphic
	model simultaneously allows to describe the onset of band-gaps in
	mechanical metamaterials together with the possibility of non-local
	effects.
	\item In section 3 we recall some results rigorously derived in \cite{madeo2016reflection}
	concerning the conservation of total energy in relaxed micromorphic
	media. The explicit form of the energy fluxes is presented both in
	the general case and using the plane wave ansatz. For completeness,
	the conservation of total energy is recalled also for classical Cauchy
	continua.
	\item In section 4 a particular connection between a Cauchy medium and a
	relaxed micromorphic medium is introduced on the basis of the results
	proposed in \cite{madeo2016reflection}. This connection has been called ``\textit{macro
		internal clamp with free microstructure}'' and allows continuity
	of macroscopic displacement at the considered interface together with
	free motions of the microstructure on the side of the interface occupied
	by the relaxed micromorphic medium. Plane wave solutions are presented
	for the displacement in the Cauchy side and for the displacement and
	the micro-distortion in the relaxed micromorphic side. The unknown
	amplitudes of the reflected and transmitted waves are calculated by
	imposing the constraint of ``\textit{macro internal clamp with free
		microstructure}'' at the Cauchy/relaxed micromorphic interface.
	The presented general study is particularized to the case
	in which only longitudinal waves travel in the considered materials.
	\item In section 5 reflection and transmission coefficients at a Cauchy/relaxed-micromorphic 	interface are defined: they measure the percentage of the energy initially carried by the incident wave which is reflected or transmitted at 	the interface. The particular degenerate limit case of 	the relaxed micromorphic model which is obtained by setting $L_{c}=0$ is then introduced and some characteristic frequencies which allow to determine the bounds of the band-gaps for longitudinal waves are defined as functions of the constitutive elastic parameters of the model. Such degenerate limit case, often referred to as \textit{internal variable model} can be used as a first rough fitting of the relaxed micromorphic model on real experimental data but it is not able to account for non-local effects. Indeed, internal variable models are able to catch the main macroscopic features of band-gap metamaterials \cite{pham2013transient,sridhar2016homogenization}. It is then argued that the fact of switching on the parameter $L_{c}$ would actually allow to perform a more refined fitting on the experimental profiles of the 	reflection coefficient.
	\item In section 6 the profile of the reflection coefficient as obtained
	by using our relaxed micromorphic model with $L_{c}=0$ is compared
	to the analogous profile presented in \cite{lucklum2012two}.
	As a result of this direct comparison with experimentally-based results, four conditions
	on the elastic parameters are established which allow for
	the determination of four constitutive parameters as function of the fifth which remains free and is calibrated in order to obtain the best fitting with the experimental profile of the reflection coefficient. Finally, the characteristic length $L_{c}$ is switched on
	and it is tuned to perform a more refined fitting until the theoretical
	profile of the reflection coefficient is more precisely superimposed to the
	experimentally-based one. The estimated value of non-local effects is found
	to be $L_{c}\simeq0.5\:mm$, which means that such characteristic
	length is almost $1/3$ of the diameter of the holes embedded in the
	considered metamaterial. Such non-local effects, which are already
	non-negligible, would become more and more important if considering
	higher contrasts of the mechanical properties at the microscopic level.
	This situation could be e.g. achieved by filling adjacent holes with
	fluids with highly contrasted properties. Only one elastic parameter,
	namely the Cosserat couple modulus $\mu_{c}$ remains undetermined
	at the end of the present work. In order to estimate its value (which
	we know to be non-vanishing for a material showing complete band-gaps \cite{madeo2014band,madeo2015wave,madeo2016complete}) for the phononic crystal considered here, we would need to extend our study for longitudinal waves to the case of transverse waves. A complete determination of the whole set of constitutive coefficients for the relaxed micromorphic model is left to a forthcoming contribution.
	 \end{itemize}

\section{Dynamic formulation of the equilibrium problem}

In the micromorphic model, the kinematics is enriched with respect to classical Cauchy continua by introducing an additional tensor  field of \textbf{non-symmetric micro-distortions} $\p:\Omega\subset\R^{3}\rightarrow\R^{3\times3}$, beyond the classical macroscopic displacement $u:\Omega\subset\R^{3}\rightarrow\R^{3}$. Then, a \textbf{non-symmetric elastic (relative) distortion} $e=\nablau-\p$ can be defined and the modeling proceeds by obtaining the constitutive relations linking elastic-distortions to stresses and by postulating a balance equation for the micro-distortion field $\p$. All such steps can be preferably done in a variational framework such that only energy contributions need to be defined a priori. For the dynamic case, one adds in the kinetic energy density so-called micro-inertia density contributions, acting on the time derivatives $\p_{,t}$ of the micro-distortion terms.

\subsection{The classical Cauchy medium}

In this subsection we recall that the strain energy density W and the kinetic energy T for a classical Cauchy medium in the isotropic setting take the form\footnote{Here and in the sequel we denote by the subscript $,t$ the partial derivative with respect to time of the considered field.}

\begin{align}
W & =\mu\left\Vert \,\mathrm{sym}\,\nablau\,\right\Vert ^{2}+\frac{\lambda}{2}\left(\mathrm{tr}\left(\mathrm{sym}\,\nablau\right)\right)^{2},\qquad\qquad T=\frac{1}{2}\rho\left\Vert  u_{,t}\right\Vert ^{2}, \label{Pot-Cauchy}
\end{align}
where $\lambda$ and $\mu$ are the classical Lamé parameters and $ u$ denotes the classical macroscopic displacement field. 

The associated bulk equations of motion in strong form, obtained by a classical least action principle, take the usual form:
\begin{align}
\rho\, u_{,tt} & =\Div\:\Sig, \qquad &\forall& x\in\Omega,\label{eq:Motion_Cauchy}\\\nonumber
 f:\hspace{-0.1cm}&=\Sig\x \n=0 \qquad \mathrm{or} \qquad u=u_{0}, \qquad &\forall& x\in\partial\Omega,
\end{align}
where $\n$ is the normal to the boundary $\partial\Omega$, and  $\Sig$ is the symmetric elastic stress tensor defined as:
\begin{align}
\Sig\left(\nablau\right)=\,2\,\mu\, \sym \nablau+\lambda\tr (\nablau)\mathds{1}\,. & \label{eq:SigmaClas}
\end{align}
Considering the case of \textbf{plane waves}, we suppose that the space dependence of all introduced kinematic fields are limited to the component $x_{1}$ of $x$ which is the direction of propagation of the wave. With this hypothesis, see\cite{madeo2016reflection}, the equations of motion \eqref{eq:Motion_Cauchy}$_{1}$ become
\begin{gather}
\underbrace{{u}_{1,tt}=\frac{\lambda+2\mu}{\rho}\,u_{1,11}}_{	\text{longitudinal}},\qquad\qquad\underbrace{{u}_{2,tt}=\frac{\mu}{\rho}\,u_{2,11}}_{	\text{transverse 2}},\qquad\qquad\underbrace{{u}_{3,tt}=\frac{\mu}{\rho}\,u_{3,11}}_{	\text{transverse 3}}.\label{CauchyAS}
\end{gather}
We now look for solutions of the dynamic problem \eqref{CauchyAS} in the form 
\begin{align}
u(x,t)=\alpha \, e^{i \left(k\, x_{1}-\,\omega \,t\right)}, \qquad\qquad\alpha\in\R^{3}.\label{WaveForm1}
\end{align} 
Considering a wave traveling in an \textbf{infinite domain} no conditions on the boundary are to be imposed and, replacing the wave form expression \eqref{WaveForm1} in the bulk equation \eqref{eq:Motion_Cauchy}, we can find the standard dispersion relations for Cauchy media (see also \cite{madeo2016reflection}) obtaining
\begin{align}
\underbrace{\omega^{2}=c_{l}^{2}k^{2}}_{	\text{longitudinal}},\qquad\qquad\underbrace{\omega^{2}=c_{t}^{2}k^{2}}_{	\text{transverse 2}},\qquad\qquad\underbrace{\omega^{2}=c_{t}^{2}k^{2}}_{	\text{transverse 3}},\label{CauchyAS-1}
\end{align} 
where we denoted by 
\begin{align}
c_{l}=\sqrt{\frac{\lambda+2\mu}{\rho}},\qquad c_{t}=\sqrt{\frac{\mu}{\rho}},
\end{align} 
the characteristic speeds in classical Cauchy media of longitudinal
and transverse waves, respectively. The dispersion relations
can be traced in the plane $(\omega,k)$, giving rise to the standard
non-dispersive behavior for a classical Cauchy continuum (see \cite{madeo2016reflection,chen2003connecting,chen2003determining,chen2004atomistic,achenbach1973wave}).
Indeed it is easily seen that for Cauchy continua the relations \eqref{CauchyAS-1} can be inversed as: 
\begin{align}
\underbrace{k=\pm \frac{1}{c_{l}}\omega}_{	\text{longitudinal}}, \qquad\qquad \underbrace{k=\pm \frac{1}{c_{t}}\omega}_{	\text{transverse 2}},\qquad\qquad \underbrace{k=\pm \frac{1}{c_{t}}\omega}_{	\text{transverse 3}}. 
\end{align}

\subsection{The relaxed micromorphic model}

Our novel relaxed micromorphic model endows Mindlin-Eringen's representation
with the second order \textbf{dislocation density tensor}  $\alpha=-\Curl \p$ instead of the full gradient $\nablap$.\footnote{The dislocation tensor is defined as $\alpha_{ij}=-\left(\Curl \p \right)_{ij}=-\p_{ih,k}\epsilon_{jkh}$, where $\epsilon$ is the Levi-Civita tensor.} In the isotropic case the energy reads

\begin{align}
W=&\underbrace{\me\,\lVert \sym\left(\nablau-\p\right)\rVert ^{2}+\frac{\lle}{2}\left(\mathrm{tr} \left(\nablau-\p\right)\right)^{2}}_{\mathrm{{\textstyle isotropic\ elastic-energy}}}	+\hspace{-0.1cm}\underbrace{\mc\,\lVert \skew\left(\nablau-\p\right)\rVert ^{2}}_{\mathrm{\textstyle rotational\   elastic\ coupling}		}\hspace{-0.1cm} \label{eq:Ener-2}\\
& \quad 
+\underbrace{\mh\,\lVert \sym \p\rVert ^{2}+\frac{\lh}{2}\,\left(\mathrm{tr} \p\right)^{2}}_{\mathrm{{\textstyle micro-self-energy}}}
+\hspace{-0.2cm}\underbrace{\frac{\mLc}{2} \,\lVert \Curl \p\rVert^2}_{\mathrm{\textstyle isotropic\ curvature}}\,,
\nonumber  
\end{align}
where the parameters and the elastic stress are analogous to the standard Mindlin-Eringen micromorphic model. The model is well-posed in the statical and dynamical case including when $\mc=0$, see \cite{neff2015relaxed,ghiba2014relaxed}.

In our relaxed model the complexity of the general micromorphic model has been decisively reduced featuring basically only symmetric strain-like variables and the $\Curl$ of the micro-distortion $\p$. However, the relaxed model is still general enough to include the full micro-stretch as well as the full Cosserat micro-polar model, see \cite{neff2014unifying}. Furthermore, well-posedness results for the statical and dynamical cases have been provided in \cite{neff2014unifying} making decisive use of recently established new coercive inequalities, generalizing Korn's inequality to incompatible tensor fields \cite{neff2015poincare,neff2002korn,neff2012maxwell,neff2011canonical,bauer2014new,bauer2016dev}.

The relaxed micromorphic model counts 6 constitutive parameters in the isotropic case ($\me$, $\lle$, $\mh$, $\lh$, $\mc$, $L_c$). The characteristic length $L_c$ is intrinsically related to non-local effects due to the fact that it weights a suitable combination of first order space derivatives in the strain energy density \eqref{eq:Ener-2}. For a general presentation of the features of the relaxed micromorphic model in the anisotropic setting, we refer to \cite{barbagallo2016transparent}.

As for the kinetic energy, we consider that it takes the following
form 
\begin{gather}
T=\frac{1}{2}\rho\left\Vert u_{,t}\right\Vert ^{2}+\frac{1}{2}\eta\left\Vert \p_{,t}\right\Vert ^{2},\label{eq:Kinetic_energy2}
\end{gather}
where $\rho$ is the value of the averaged macroscopic mass density of the considered metamaterial, while $\eta$ is the micro-inertia density. In the following numerical simulations we will suppose that the macroscopic density $\rho$ is known, while we will deduce the micro-inertia parameter $\eta$ by an inverse approach. In any case, we have checked that the value of $\rho$ does not influence the profile of the reflection coefficient, at least for the considered range of frequencies.

Defining the elastic stress $\sig$, the hyper-stress tensor $ m$, the micro-stress $ s$ as: 
\begin{align}
\sig\left(\nablau,\p\right)&=\,2\,\me\, \sym\left(\nablau-\p\right)+2\,\mc \,\skew\left(\nablau-\p\right)+\lle\tr\left(\nablau-\p\right)\mathds{1}\,,
\label{eq:Constitutive_relaxed}\\
 m\left(\Curl\,\p\right)&= \mLc\,\Curl\,\p, \qquad  \qquad  s(\p)=\left[2\mh \sym \p +\lh \tr (\p) \mathds{1}\right],
\nonumber 
\end{align}
the associated equations of motion in strong form, obtained by a classical least action principle take the form (see \cite{madeo2016reflection,madeo2015wave,neff2015relaxed,madeo2014band})
\begin{align}
\rho\,  u_{,tt}&=\,\Div\,\sig, \hspace{-1cm} & &&&  \forall x\in\Omega,\label{eq:Dyn}\\\nonumber
\eta\,  \p_{,tt}&=\sig - s- \Curl\, m ,\hspace{-1cm} & & &  &\forall x\in\Omega,
\end{align}
while the associated natural and kinematical boundary condition are\footnote{Here and in the sequel a central dot indicates a simple contraction between tensors of suitable order. For example, $\left(A\cdot v\right)_i=A_{ij}v_j$ and $\left(A\cdot B\right)_{ij}=A_{ih}B_{hj}$. Einstein convention of sum of repeated indexes is used throughout the paper}:
\begin{align}
 t&:=\sig\x \n=0 \hspace{-1cm} &\mathrm{or} \qquad\quad&\,u=u_{0},& \qquad &\forall x\in\partial\Omega,\label{eq:Dyn2}\\
\tau&:= m\x\epsilon\x\n=0 \hspace{-1cm} &\mathrm{or}  \qquad\quad&\p\x \nu_{1}=p_{1}, \quad \p\x \nu_{2}=p_{2}, & \qquad &\forall x\in\partial\Omega,\nonumber
\end{align}
where $\n$ and $\nu_{i}$ ($i=1,2$) are the normal and tangent vectors to the boundary $\partial\Omega$, while $ t$ and $\tau$ are the resulting internal force and double force vectors.

Our approach consists in writing the micro-distortion tensor $\p\in\R^{3\times3}$ by means of its Cartan-Lie decomposition as $\p=\devsym\left(\p\right)+\skew\left(\p\right)+\frac{1}{3}\tr\left( \p\right)\id$. Therefore, defining
\begin{align}
\p_{\left[ij\right]}&:=\left(\skew \p \right)_{ij}= \frac{1}{2}\left(\p_{ij}-\p_{ji}\right),&\p^{S}& := \frac{1}{3}\tr\left( \p\right), \\\nonumber
\p_{\left(ij\right)}
&:= \left(\sym\p \right)_{ij}=\frac{1}{2}\left(\p_{ij}+\p_{ji}\right),&\p_{\left(ij\right)}^D
&:= \left(\devsym\p \right)_{ij}=\p_{\left(ij\right)}-\p^{S} \delta_{ij}.   
\end{align}
we have:
\begin{align}
\p_{ij}=\p^D_{\left(ij\right)}+\p_{\left[ij\right]}+\p^{S}\delta_{ij}=\p_{\left(ij\right)}+\p_{\left[ij\right]}.
\end{align}
Furthermore, we define
\begin{align}
\p^{D}:=\p^D_{\left(11\right)}=\p_{\left(11\right)}- \p^S, \qquad  \qquad \p^{V}:=\p_{\left(22\right)}-\p_{\left(33\right)}=\p_{22}-\p_{33}.
\end{align}
Since it will be useful in the following, let us collect some of the new variables of our problem as 
\begin{equation}
 v_{1}:=\left(u_{1},P^{D},P^{S}\right),\qquad\qquad v_{2}:=\left(u_{2},P_{(12)},P_{[12]}\right),\qquad\qquad v_{3}:=\left(u_{3},P_{(13)},P_{[13]}\right),\label{eq:Unknowns1}
\end{equation}
and, for having a homogeneous notation, let us set 
\begin{equation}
v_{4}:=P_{\left(23\right)},\qquad\qquad v_{5}:=P_{[23]},\quad\qquad v_{6}:=P^{V}.\label{eq:Unknowns2}
\end{equation}
with these new variables the equations of motion \eqref{eq:Dyn} can be written as (see also \cite{madeo2016reflection}):
\begin{align}
&\underbrace{v_{1,tt}  =A\,_{1}^{R}\cdot v''_{1}+ B\,_{1}^{R}\cdot v_{1}'+ C\,_{1}^{R}\cdot v_{1}}_{\text{Longitudinal}}, 
&\underbrace{v_{\alpha,tt}= A_{\alpha}^{R}\cdot v''_{\alpha}+ B\,_{\alpha}^{R}\cdot v'_{\alpha}+ C\,_{\alpha}^{R}\cdot v_{\alpha}}_{\text{Transverse $\alpha$}},
	&\quad\ \alpha=2,3\vspace{1.2mm}\nonumber
	\\
	\label{eq:BulkEq}
	\\
&\underbrace{v_{4,tt}  =A_{4}^{R} v_{4}''+C_{4}^{R} v_{4}}_{\text{Uncoupled}}, 
	&\underbrace{v_{5,tt}=A_{5}^{R} v_{5}''+C_{5}^{R} v_{5}}_{\text{Uncoupled}},\qquad\qquad\qquad\  
	&	\underbrace{v_{6,tt}=A_{6}^{R} v_{6}''+C_{6}^{R} v_{6}}_{\text{Uncoupled}},\nonumber 
\end{align}
where, from now on, $\left(\cdot\right)'$ denotes the derivative of the quantity
$\left(\cdot\right)$ with respect to $x_{1}$ and we set 
\begin{align}
 \ A\,_{1}^{R}&=\begin{pmatrix}\frac{\lle+2\me}{\rho} & 0 & 0\vspace{1.2mm}\\
0 & \frac{\me\,L_{c}^{2}}{3\eta} & -\frac{2\,\me\,L_{c}^{2}}{3\eta}\vspace{1.2mm}\\
0 & -\frac{\me\,L_{c}^{2}}{3\eta} & \frac{2\,\me\,L_{c}^{2}}{3\eta}
\end{pmatrix},\qquad &A\,_{\alpha}^{R}&=\begin{pmatrix}\frac{\me+\mu_{c}}{\rho} & 0 & 0\vspace{1.2mm}\\
0 & \frac{\me\,L_{c}^{2}}{2\eta} & \frac{\me\,L_{c}^{2}}{2\eta}\vspace{1.2mm}\\
0 & \frac{\me\,L_{c}^{2}}{2\eta} & \frac{\me\,L_{c}^{2}}{2\eta}
\end{pmatrix}, &\alpha&=2,3,\vspace{1.2mm}\nonumber \\
  B\,_{1}^{R}&=\begin{pmatrix}0 & -\frac{2\me}{\rho} & -\frac{3\lle+2\me}{\rho}\vspace{1.2mm}\\
\frac{4}{3}\,\frac{\me}{\eta} & 0 & 0\vspace{1.2mm}\\
\frac{3\lle+2\me}{3\eta} & 0 & 0
\end{pmatrix},\qquad &B\,_{\alpha}^{R}&=\begin{pmatrix}0 & -\frac{2\me}{\rho} & \frac{2\mu_{c}}{\rho}\vspace{1.2mm}\\
\frac{\me}{\eta} & 0 & 0\vspace{1.2mm}\\
-\frac{\mu_{c}}{\eta} & 0 & 0
\end{pmatrix}, &\alpha&=2,3,\vspace{1.2mm}\label{eq:DefMatr}\\
C\,_{1}^{R}&=\begin{pmatrix}0 & 0 & 0\vspace{1.2mm}\\
0 & -\frac{2\left(\me+\mh\right)}{\eta} & 0\vspace{1.2mm}\\
0 & 0 & -\frac{\left(3\lle+2\me\right)+\left(3\lh+2\mh\right)}{\eta}
\end{pmatrix}, &C\,_{\alpha}^{R}&=\begin{pmatrix}0 & 0 & 0\vspace{1.2mm}\\
0 & -\frac{2\left(\me+\mh\right)}{\eta} & 0\vspace{1.2mm}\\
0 & 0 & -\frac{2\mu_{c}}{\eta}
\end{pmatrix},\ &\alpha&=2,3,\vspace{1.2mm}\nonumber \\
A_{4}^{R}&=A_{5}^{R}=A_{6}^{R}=\frac{\me\,L_{c}^{2}}{\eta},\qquad\quad\ C_{5}^{R}=-\frac{2\mu_{c}}{\eta}, &C_{4}^{R}&=C_{6}^{R}=-\frac{2\left(\me+\mh\right)}{\eta}.\nonumber 
\end{align}

As done for the Cauchy continuum, we consider the case of \textbf{plane waves}, by looking for solutions of the dynamic problem in the form
\begin{align}
v_{1}=\beta_{1}\,e^{i(kx_{1}-\omega \, t)},&\qquad & v_{2}=\beta_{2}\,e^{i(kx_{1}-\omega \, t)},&\qquad & v_{3}=\beta_{3}\,e^{i(kx_{1}-\omega \, t)},
\label{eq:WaveForm}\\
v_{4}=\beta_{4}\,e^{i(kx_{1}-\omega \, t)},&\qquad & v_{5}=\beta_{5}\,e^{i(kx_{1}-\omega \, t)},&\qquad & v_{6}=\beta_{6}\,e^{i(kx_{1}-\omega \, t)}.\nonumber 
\end{align}
where $\beta_{1}, \beta_{2}, \beta_{3} \in \R^{3}$ and $\beta_{4}, \beta_{5}, \beta_{6} \in \R$. Replacing this wave-form in equations \eqref{eq:BulkEq} the dispersion relations for the relaxed micromorphic medium can be obtained (see \cite{madeo2016reflection} for details).

It is clear that the study of dispersion relations for the relaxed micromorphic continuum is intrinsically more complicated than in the case of classical Cauchy continuum due to its enriched kinematics. In the papers \cite{madeo2016reflection,madeo2015wave,madeo2014band} it was explicitly pointed out that the wave-numbers for uncoupled waves in the relaxed micromorphic continua can be calculated as function of the frequency $\omega$ as:
\begin{align}
\underbrace{k=\pm \frac{1}{c_{m}}\sqrt{\omega^2-\omega_{s}^2}}_{\mathrm{Uncoupled \ v_4}},\qquad\qquad\qquad 
\underbrace{k=\pm \frac{1}{c_{m}}\sqrt{\omega^2-\omega_{r}^2}}_{\mathrm{Uncoupled \ v_5}},\qquad\qquad\qquad 
\underbrace{k=\pm \frac{1}{c_{m}}\sqrt{\omega^2-\omega_{s}^2}}_{\mathrm{Uncoupled \ v_6}},
\end{align}   
where we set
\begin{align}
\omega_{s}=\sqrt{\frac{2\left(\me+\mh\right)}{\eta}},\qquad\qquad\qquad\quad \omega_{r}=\sqrt{\frac{2\mc}{\eta}},\qquad\qquad\qquad\quad c_{m}=\sqrt{\frac{\mLc}{\eta}}.\qquad\qquad
\end{align}
These relationships can be eventually easily inversed to find $\omega$ as function of $k$.

As far as longitudinal and transverse waves are concerned, the expressions for the wave-numbers $k$ which allow for non-trivial solutions are by far more complicated. We refer to \cite{madeo2016reflection} for the complete set-up of the eigenvalue problems which must be solved to find the explicit expressions for the wave-numbers, limiting ourselves here to denote them by $\pm k_{1}^{1},\,\pm k_{1}^{2}$ for longitudinal waves and $\pm k_{\alpha}^{1},\,\pm k_{\alpha}^{2},\, \alpha=2,3$ for transverse waves. Of course, the computed expressions for $k_{1}^{1},\, k_{1}^{2},\, k_{\alpha}^{1},\, k_{\alpha}^{2}$ depend on the elastic constitutive parameters of the relaxed micromorphic model appearing in \eqref{eq:Ener-2} and on the frequency $\omega$. 

\begin{figure}[H]
	\begin{centering}
		\begin{tabular}{ccccc}
			\includegraphics[width=5cm]{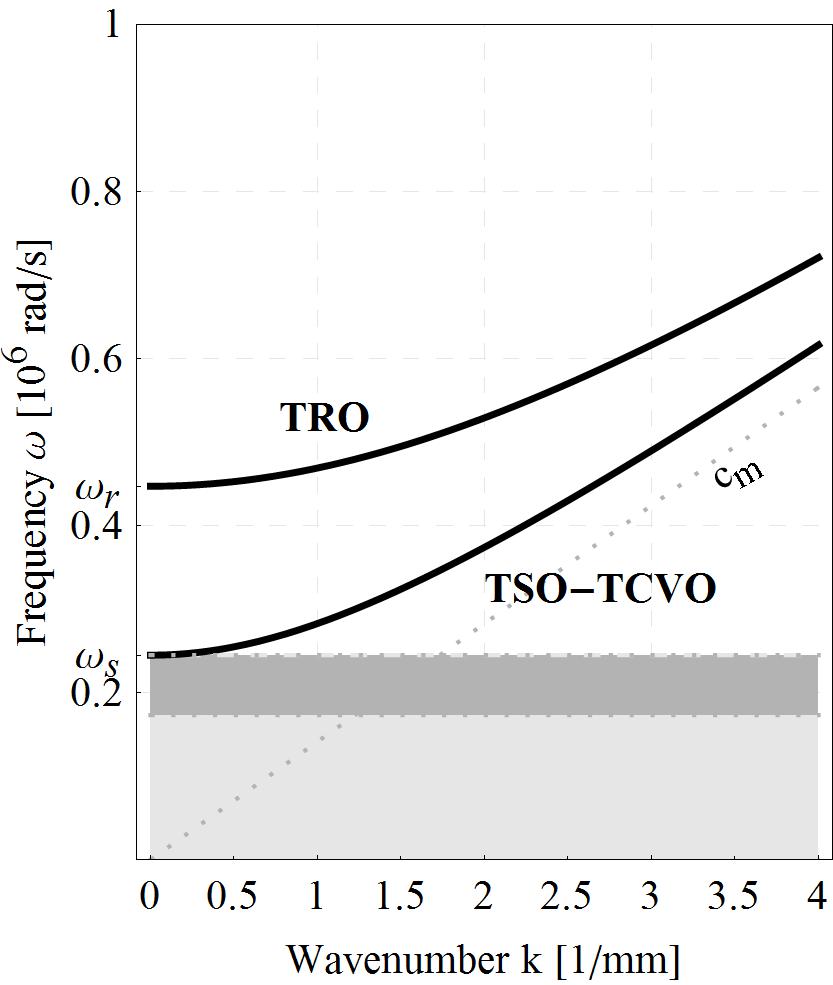}  &
			\includegraphics[width=5cm]{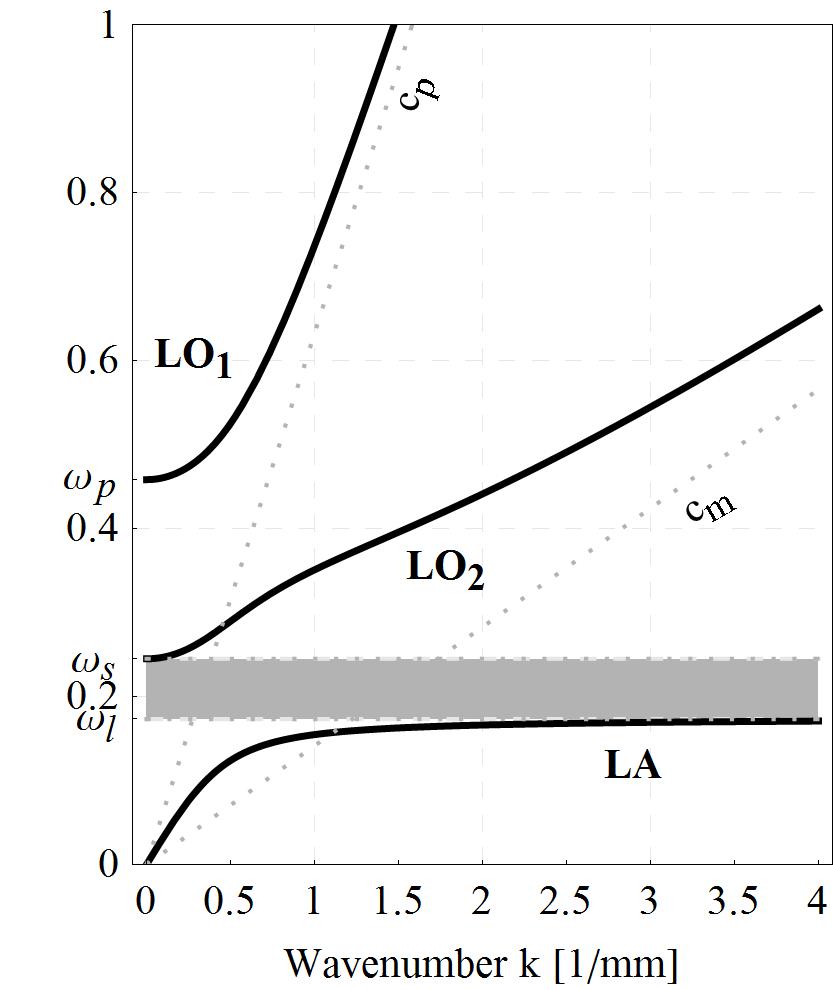} &
			\includegraphics[width=5cm]{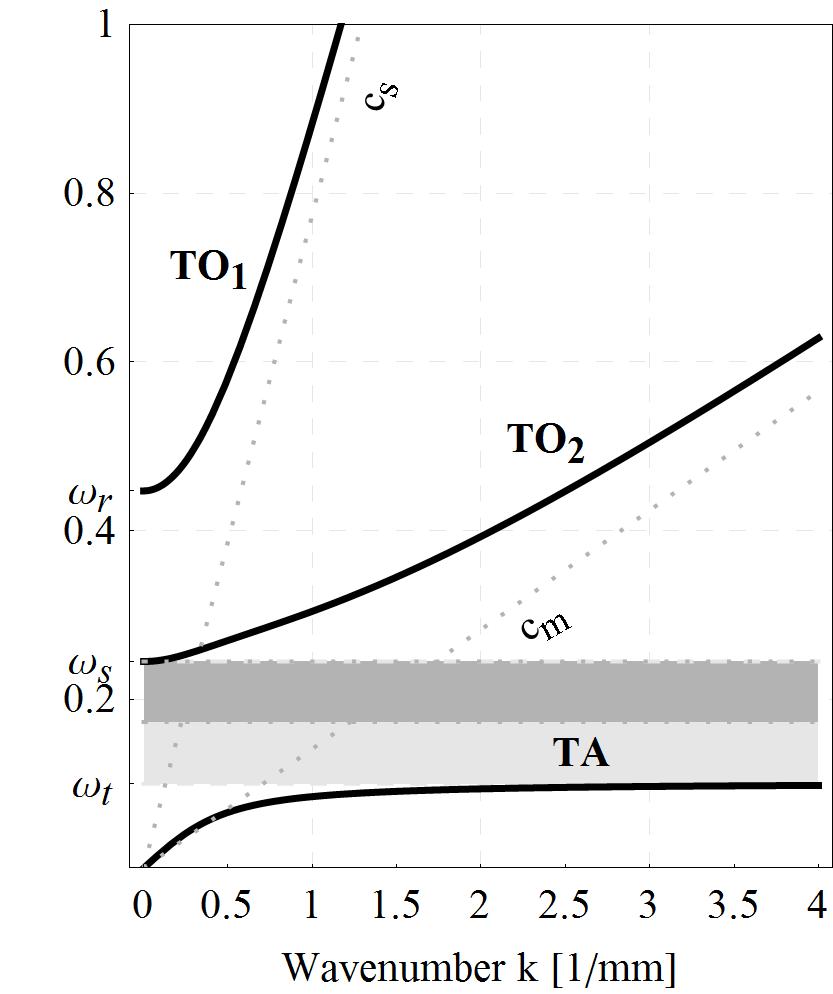} \\
			\quad(a)&\quad(b)&\quad(c)
		\end{tabular}
		\par\end{centering}
	
	\caption{\label{CurlNon}Dispersion relations $\omega=\omega(k)$ for the
		\textbf{relaxed micromorphic model}  with non-vanishing Cosserat couple modulus $\mc>0$.  Uncoupled waves (a), longitudinal waves (b) and
		transverse waves (c). TRO: transverse rotational optic, TSO: transverse shear optic, TCVO:
		transverse constant-volume optic, LA: longitudinal acoustic, LO$_{1}$-LO$_{2}$:
		$1^{st}$ and $2^{nd}$ longitudinal optic, TA: transverse acoustic, TO$_{1}$-TO$_{2}$: $1^{st}$ and $2^{nd}$ transverse optic.
	}	
\end{figure}

We present here the \textbf{dispersion relations} for longitudinal, transverse and uncoupled waves obtained with a non vanishing Cosserat modulus $\mc>0$ (Figure \ref{CurlNon}). A \textbf{complete frequency band-gap} can be recognized in the shaded intersected domain bounded from the maximum between $\omega_{l}=\sqrt{\frac{\mh+2\,\lh}{\eta}}$ and $\omega_{t}=\sqrt{\frac{\mh}{\eta}}$ and the minimum between $\omega_{r}$ and $\omega_{s}$. The existence of the complete band gap is related to $\mc>0$ via the cut-off frequency $\omega_{r}=\sqrt{\frac{2\mc}{\eta}}$ of the uncoupled waves TRO and the transverse wave TO1.

The relaxed micromorphic model is the only generalized continuum model (see \cite{madeo2016complete,madeo2016reflection}) which is able to simultaneously account for:
\begin{itemize}
	\item complete frequency band-gaps
	\item non-local effects involving interactions between adjacent unit cells.
\end{itemize}
Non locality is an intrinsic feature of metamaterials with heterogeneous microstructure and it is sensible that it plays a non-negligible role on the phenomena of wave propagation, reflection and transmission. We will show in the remainder of this paper, based on real experiments, that the characteristic length associated to such non-local effects may be an order of magnitude comparable with the characteristic size of the underlying microstructure, even if considering otherwise homogeneous metamaterials. 

Indeed, when considering metamaterials with heterogeneous microstructures (e.g. metamaterials in which two adjacent unitary cells have strongly contrasted mechanical properties) the non-local effects may play a more and more important role on the overall behavior of the considered metamaterial.

\section{Conservation of the total energy}

It is known that if one considers conservative mechanical systems,
like in the present paper, then conservation of total energy must
be verified in the form 
\begin{equation}
\frac{dE}{dt}+\mathrm{div}\: H=0,\label{EnnergyConservation}
\end{equation}
where $E=T+W$ is the total energy of the considered system and $ H$ is the energy flux vector. It is clear that the explicit expressions for the total energy and for the energy flux are different depending on whether one considers a classical Cauchy model or a relaxed micromorphic one. If the expression of the total energy $E$ is straightforward for the two mentioned cases (it suffices to look at the given expressions of $T$ and $W$), the explicit expression of the energy flux $ H$ is more complicated to be obtained. The explicit expression of the energy fluxes for the Cauchy and relaxed micromorphic media have been deduced in \cite{madeo2016reflection}, to which we refer for additional details on this subject.

\subsection{The classical Cauchy medium}
In classical Cauchy continua the energy flux vector $ H$ can be
written as
\begin{equation}
 H=-\Sig\cdot u_{,t},\label{Flux-Cauchy}
\end{equation}
where the symmetric Cauchy stress tensor $\Sig$ has been defined in equation \eqref{eq:SigmaClas} in terms of the displacement field. The first component of the energy flux vector given in equation \eqref{Flux-Cauchy},
simplifies in the 1D case into
\begin{equation}
H_{1}=-\dot{u}_{1}\left[\left(\lambda+2\mu\right)\:u_{1,1}\right]-\dot{u}_{2}\left[\mu\,u_{2,1}\right]-\dot{u}_{3}\left[\mu\,u_{3,1}\right].\label{Flux-Cauchy-1}
\end{equation}

\subsection{The relaxed micromorphic continuum}
In relaxed micromorphic media, the energy flux vector $H$ is defined as (see \cite{madeo2016reflection})\footnote{The symbol $:$ indicates the double contraction between tensors of suitable order. For example,  $\left(A : B\right)=A_{ij}B_{ji}$ and $\left(C : B\right)_{i}=C_{ijh}B_{hj}$.}

\begin{equation}
H=- \sig^{T}\cdot u_{,t}-( m^{T}\cdot\p_{,t}):\epsilon,\label{Flux}
\end{equation}
where the stress tensor $\sig$ and the hyper-stress
tensor $ m$ have been defined in equation \eqref{eq:Constitutive_relaxed}
in terms of the basic kinematical fields and $\epsilon$ is the Levi-Civita tensor.

When considering conservation of total energy, it can be checked that
the first component of the energy flux \eqref{Flux} can be rewritten
in terms of the new variables as
\begin{equation}
\widetilde{H}_{1}=H_{1}^{1}+H_{1}^{2}+H_{1}^{3}+H_{1}^{4}+H_{1}^{5}+H_{1}^{6}\label{eq:FluxRelaxed}
\end{equation}
with
\begin{align}
H_{1}^{1}&=\dot{ v}_{1}\cdot\left[\left(\begin{array}{ccc}
-\left(\lle+2\me\right) & 0 & 0\\
0 & -\frac{\mLc}{2} & \mLc\\
0 & \mLc & -2\mLc
\end{array}\right)\text{\ensuremath{\cdot}} v_{1}'+\left(\begin{array}{ccc}
0 & 2\me & \left(3\lle+2\me\right)\\
0 & 0 & 0\\
0 & 0 & 0
\end{array}\right)\text{\ensuremath{\cdot}} v_{1}\right],\nonumber \\ \nonumber \\ 
H_{1}^{2}&=\dot{ v}_{2}\cdot\left[\left(\begin{array}{ccc}
-\left(\me+\mu_{c}\right) & 0 & 0\\
0 & -\mLc & -\mLc\\
0 & -\mLc & -\mLc
\end{array}\right)\text{\ensuremath{\cdot}} v_{2}'+\left(\begin{array}{ccc}
0 & 2\me & -2\text{\ensuremath{\mu}}_{c}\\
0 & 0 & 0\\
0 & 0 & 0
\end{array}\right)\text{\ensuremath{\cdot}} v_{2}\right],
\nonumber \\ \label{Fluxes}\\
H_{1}^{3}&=\dot{ v}_{3}\cdot\left[\left(\begin{array}{ccc}
-\left(\me+\mu_{c}\right) & 0 & 0\\
0 & -\mLc & -\mLc\\
0 & -\mLc & -\mLc
\end{array}\right)\text{\ensuremath{\cdot}} v_{3}'+\left(\begin{array}{ccc}
0 & 2\me & -2\text{\ensuremath{\mu}}_{c}\\
0 & 0 & 0\\
0 & 0 & 0
\end{array}\right)\text{\ensuremath{\cdot}} v_{3}\right],\nonumber \\\nonumber \\ 
H_{1}^{4}&=-2\mLc\left(\mathrm{v}_{4}\right)_{,1}\dot{\mathrm{v}}_{4},\qquad H_{1}^{5}=-2\mLc\left(\mathrm{v}_{5}\right)_{,1}\dot{\mathrm{v}}_{5},\qquad H_{1}^{6}=-\frac{\mLc}{2}\left(\mathrm{v}_{6}\right)_{,1}\dot{\mathrm{v}}_{6}.\nonumber 
\end{align}

\section{Interface jump conditions at a Cauchy/relaxed-micromorphic interface}
In this section we present a possible choice of boundary conditions to be imposed between a Cauchy medium and a relaxed micromorphic medium. Such set of boundary conditions has been derived in \cite{madeo2016reflection} and allows to describe free vibrations of the microstructure at the considered interface. We will show in the remainder of this paper how this particular choice of the boundary conditions is capable to describe phenomena of wave transmission in real mechanical metamaterials. For the full presentation of the complete sets of possible connections that can be established at Cauchy/relaxed, relaxed/relaxed, Cauchy/Mindlin, Mindlin/Mindlin interfaces we refer to\cite{madeo2016reflection}.

When considering connections between a Cauchy and a relaxed micromorphic medium one can impose more kinematical boundary conditions than in the case of connections between Cauchy continua. More precisely, one can act on the displacement field $\text{\ensuremath{ u}}$ (on both sides of the interface) and also on the tangential micro-distortion
$\text{\ensuremath{\p}}$ (on the side of the interface occupied
by the relaxed micromorphic continuum).  In what follows, we consider
the ``-'' region occupied by the Cauchy continuum and the ``+''
region occupied by the micromorphic continuum, so that, accordingly,
we use the following notations:
\begin{align}
	 f & =  \Sig^{-}\cdot \n^{-},\qquad  t=\sig^{+}\cdot \n^{+},\qquad\tau= \mLc\,(	\Curl\,\p^{+}) \x \epsilon \x \n^{+}.
\end{align}
It is possible to check that considering the normal $\n=(1,0,0)$, the normal components $\tau_{11},\tau_{21}$ and $\tau_{31}$ of the double force are identically zero. Therefore, the number of independent conditions that one can impose on the micro-distortions is 6 when considering a relaxed micromorphic model.

In this paper we focus our attention on one particular type of connection between a classical Cauchy continuum and a relaxed micromorphic one, which is sensible to reproduce the real situation in which the microstructure of the band-gap metamaterial is free to vibrate independently of the macroscopic matrix. Such particular connection guarantees continuity of the macroscopic displacement and free motion of the microstructure (which means vanishing double force) at the interface:
\begin{align}
\left[[ u\right]] =0,\qquad\qquad  t- f=0,\qquad\qquad\tau\cdot\nu_{1}=\tau\cdot\nu_{2}=0.
\end{align}
We explicitly remark that continuity of displacement implies continuity of the internal forces and that the conditions on the arbitrariness of micro-motions is assured by imposing that the tangent part of the double force is vanishing. 

\begin{figure}[H]
	\begin{centering}
			
			\begin{picture}(288,160)
			\put(0,10){\line(0,1){144}}
			\put(0,10){\line(1,0){288}}
			\put(0,154){\line(1,0){288}}
			\put(144,10){\line(0,1){144}}
			\put(288,10){\line(0,1){144}}
			\multiput(151,16)(12,0){12}{
				\multiput(0,0)(0,12){12}{\circle{7.2}}}
		
		\put(30,160){Cauchy medium $\ominus$}		
		\put(146,160){relaxed micromorphic medium $\oplus$}
		

			\end{picture}
		\par\end{centering}
	
	\protect\caption{Schematics of a macro internal clamp with free microstructure at a Cauchy/relaxed-micromorphic interface.}
\end{figure}
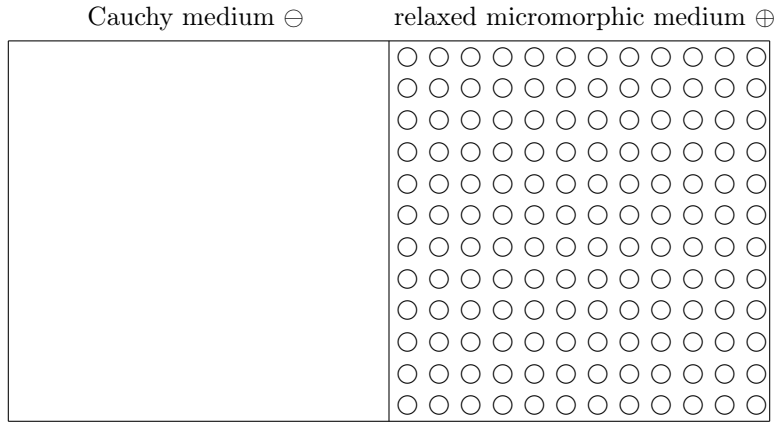

Introducing the tangent vectors $\nu_{1}=(0,1,0)$ and $\nu_{2}=(0,0,1)$ and considering the new variables presented in \eqref{eq:Unknowns1} and \eqref{eq:Unknowns2}, the boundary conditions on the jump of displacement read:
\begin{align}
 v_{1}^{+}\cdot\n-u_{1}^{-}=0,\qquad\qquad v_{2}^{+}\cdot\n-u_{2}^{-}=0,\qquad\qquad v_{3}^{+}\cdot\n-u_{3}^{-}=0,\label{Jump1}
\end{align}
while the conditions on the internal forces become (see also \cite{madeo2016reflection}):
\begin{align}
\left(\begin{array}{c}
\lle+2\me\\
0\\ 0 \end{array}\right)\cdot\left( v_{1}^{+}\right)'+\left(\begin{array}{c}
0\\ -2\me\\ -(3\lle+2\me) \end{array}\right)\cdot v_{1}^{+}&=(\lambda+2\mu)\,\left(u_{1}^{-}\right)',\nonumber 
\\\nonumber \\ 
\left(\begin{array}{c} \me+\mu_{c}\\ 0\\ 0 \end{array}\right)\cdot\left( v_{2}^{+}\right)'+\left(\begin{array}{c} 0\\ -2\me\\ 2\mu_{c}
\end{array}\right)\cdot v_{2}^{+}&=\mu\,\left(u_{2}^{-}\right)',
\label{Jump2}\\\nonumber \\ 
\nonumber \left(\begin{array}{c} \me+\mu_{c}\\ 0\\ 0 \end{array}\right)\cdot\left( v_{3}^{+}\right)'+\left(\begin{array}{c} 0\\ -2\me\\ 2\mu_{c} \end{array}\right)\cdot v_{3}^{+}&=\mu\,\left(u_{3}^{-}\right)'.
\end{align}
The conditions on the tangent part of the double force $\tau$  can be written as
\begin{align}
\tau_{22}&=\left(\begin{array}{c}
0\\ -\mLc/2\\ \mLc \end{array}\right)\cdot\left( v_{1}^{+}\right)'+\frac{\mLc}{2}\left(\mathrm{v}_{6}^{+}\right)'=0, 
&\tau_{33}&=\left(\begin{array}{c} 0\\ -\mLc/2\\ \mLc \end{array}\right)\cdot\left( v_{1}^{+}\right)'-\frac{\mLc}{2}\left(\mathrm{v}_{6}^{+}\right)'=0,
\nonumber \\\nonumber \\ 
\tau_{12}&=\left(\begin{array}{c} 0\\ \mLc\\ \mLc \end{array}\right)\cdot\left( v_{2}^{+}\right)'=0, &
\tau_{13}&=\left(\begin{array}{c} 0\\ \mLc\\ \mLc \end{array}\right)\cdot\left( v_{3}^{+}\right)'=0,
\label{Jump3}\\\nonumber \\ 
\tau_{23}&=\mLc\left(\left(\mathrm{v}_{4}^{+}\right)'+\left(\mathrm{v}_{5}^{+}\right)'\right)=0,
&\tau_{32}&=\mLc\left(\left(\mathrm{v}_{4}^{+}\right)'-\left(\mathrm{v}_{5}^{+}\right)'\right)=0,\nonumber 
\end{align}
while we recall once again that the normal part of the double force is identically vanishing, i.e.:
\begin{align}
\tau_{11}=0,\qquad\qquad
\tau_{21}=0,\qquad\qquad
\tau_{31}=0.
\end{align}

\subsection{Reflection and transmission of plane waves at a Cauchy/relaxed-micromorphic interface}

\begin{figure}[H]
	\begin{center}
		\begin{picture}(400,220)
		
		\put(18,184){\color{red}{incident longitudinal wave}}		
		\put(20,180){\color{red}{\vector(1,0){110}}}
		\put(28,168){\color{red}{$u_1^i=\overline{\alpha}_1 \,e^{i(\omega/c_{l}\,x_{1}-\omega \, t)} $}}

		\put(20,139){\color{blue}{incident transverse waves}}				
		\put(20,135){\color{blue}{\vector(1,0){110}}}
		\put(28,123){\color{blue}{$u_\alpha^i=\overline{\alpha}_\alpha \,e^{i(\omega/c_{t}\,x_{1}-\omega \, t)} $}}
		
		\put(18,94){\color{red}{reflected longitudinal wave}}				
		\put(130,90){\color{red}{\vector(-1,0){110}}}
		\put(28,78){\color{red}{$u_1^r=\alpha_1 \,e^{i(-\omega/c_{l}\,x_{1}-\omega \, t)} $}}
		
		\put(20,49){\color{blue}{reflected transverse waves}}		
		\put(130,45){\color{blue}{\vector(-1,0){110}}}	
		\put(28,32){\color{blue}{$u_\alpha^r=\alpha_\alpha \,e^{i(-\omega/c_{t}\,x_{1}-\omega \, t)} $}}

		\put(200,187){\color{red}{transmitted longitudinal waves}}
		\put(170,185){\color{red}{\vector(1,0){210}}}
		\put(180,173){\color{red}{$v_1=\beta_{1}^{1} h_{1}^{1}e^{i(k_{1}^{1}(\omega)x_{1}-\omega \, t)}+\beta_{1}^{2} h_{1}^{2}e^{i(k_{1}^{2}(\omega)x_{1}-\omega \, t)}$}}
		
		\put(200,152){\color{blue}{transmitted transverse waves}}
		\put(170,150){\color{blue}{\vector(1,0){210}}}
		\put(180,138){\color{blue}{$v_{\alpha}=\beta_{\alpha}^{1} h_{\alpha}^{1}e^{i(k_{\alpha}^{1}(\omega)x_{1}-\omega \, t)}+\beta_{\alpha}^{2} h_{\alpha}^{2}e^{i(k_{\alpha}^{2}(\omega)x_{1}-\omega \, t)}$}}
	
		\put(200,117){\color{Green}{transmitted uncoupled wave}}
		\put(170,115){\color{Green}{\vector(1,0){210}}}
		\put(210,102){\color{Green}{$	v_{4}^{+}=\beta_{4}\,e^{i(1/c_{m}\:\sqrt{\omega^{2}-\omega_{s}^{2}}\:x_{1}-\omega \, t)}$}}

		\put(200,82){\color{Green}{transmitted uncoupled wave}}		
		\put(170,80){\color{Green}{\vector(1,0){210}}}
		\put(210,67){\color{Green}{$	v_{5}^{+}=\beta_{5}\,e^{i(1/c_{m}\:\sqrt{\omega^{2}-\omega_{r}^{2}}\:x_{1}-\omega \, t)}$}}
		
		\put(200,47){\color{Green}{transmitted uncoupled wave}}	
		\put(170,45){\color{Green}{\vector(1,0){210}}}
		\put(210,32){\color{Green}{$v_{6}^{+}=\beta_{6}\,e^{i(1/c_{m}\:\sqrt{\omega^{2}-\omega_{s}^{2}}\:x_{1}-\omega \, t)}$}}
		
		\put(133,0){$x_1=0$}
		\put(360,0){$\alpha=2,3$}
		\put(40,210){Cauchy medium $\ominus$}		
		\put(200,210){relaxed micromorphic medium $\oplus$}
		
		\put(0,15){\line(0,1){190}}
		\put(150,15){\line(0,1){190}}
		\put(400,15){\line(0,1){190}}
		\put(0,15){\line(1,0){400}}
		\put(0,205){\line(1,0){400}}
		\end{picture}
	\end{center}
	\caption{\label{FigWaves} Incident, reflected and transmitted waves at a Cauchy/relaxed-micromorphic interface.}
\end{figure}
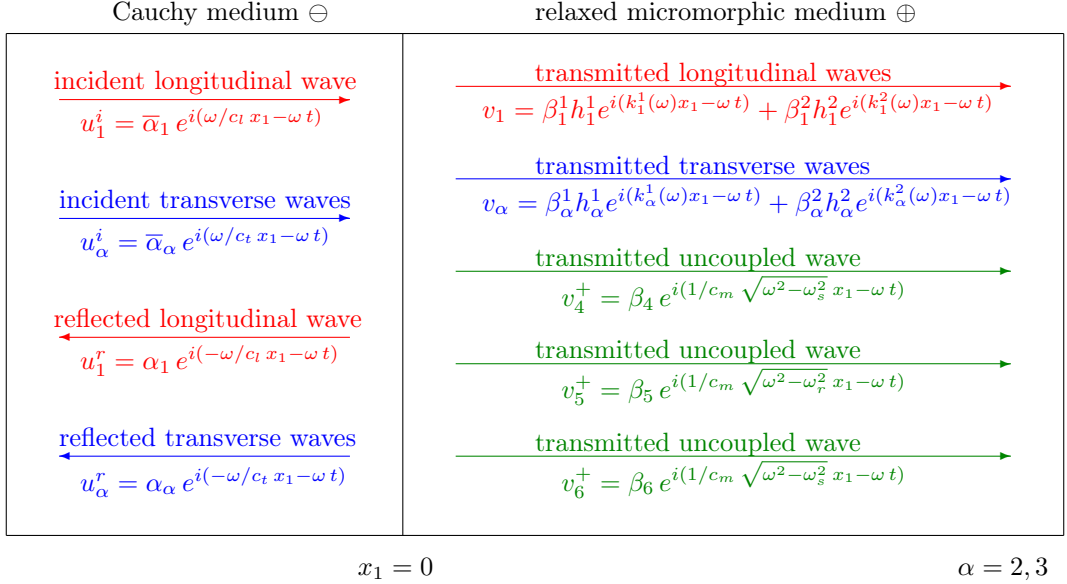

When studying the reflection and transmission of a plane wave at a Cauchy/relaxed-micromorphic interface, we are considering that an incident wave traveling in the Cauchy medium impacts the interface. Two waves are than generated, namely one wave reflected in the Cauchy medium and one transmitted in the relaxed micromorphic medium. We explicitly remark that the reflected wave contains the longitudinal and transverse parts of the displacement field, while the transmitted wave contains a longitudinal part on the field $v_1$, two transverse parts on $v_{\alpha}$ ($\alpha=2,3$) and the three uncoupled fields $v_4$, $v_5$ and $v_6$ (see also Figure \ref{FigWaves}).

Considering the wave forms \eqref{WaveForm1} and \eqref{eq:WaveForm} for the unknown fields $u^{-},\,v_{1}^{+},\,v_{\alpha}^{+},\,v_{4}^{+},\,v_{5}^{+}$ and $,\,v_{6}^{+}$ the solution of the considered problem can be written as:
\begin{align}
u_{1}^{-}&=\overline{\alpha}_{1}\,e^{i(\omega/c_{l}\,x_{1}-\omega \, t)}+\alpha_{1}\,e^{i(-\omega/c_{l}\,x_{1}-\omega \, t)},\vspace{1.2mm}
\nonumber\\
u_{2}^{-}&=\overline{\alpha}_{2}\,e^{i(\omega/c_{t}\,x_{1}-\omega \, t)}+\alpha_{2}\,e^{i(-\omega/c_{t}\,x_{1}-\omega \, t)},\vspace{1.2mm} \label{CauchyWaves}
\\\nonumber
u_{3}^{-}&=\overline{\alpha}_{3}\,e^{i(\omega/c_{t}\,x_{1}-\omega \, t)}+\alpha_{3}\,e^{i(-\omega/c_{t}\,x_{1}-\omega \, t)}
\end{align}
and moreover 
\begin{align}
v_{1}^{+}&=\beta_{1}^{1} h_{1}^{1}e^{i(k_{1}^{1}(\omega)x_{1}-\omega \, t)}+\beta_{1}^{2} h_{1}^{2}e^{i(k_{1}^{2}(\omega)x_{1}-\omega \, t)},
\quad
v_{\alpha}^{+}=\beta_{\alpha}^{1} h_{\alpha}^{1}e^{i(k_{\alpha}^{1}(\omega)x_{1}-\omega \, t)}+\beta_{\alpha}^{2} h_{\alpha}^{2}e^{i(k_{\alpha}^{2}(\omega)x_{1}-\omega \, t)},
\quad
\alpha=2,3,\vspace{1.2mm}\nonumber \\
\vspace{1.2mm}\label{eq:WaveForm-1}\\
v_{4}^{+}&=\beta_{4}\,e^{i(1/c_{m}\:\sqrt{\omega^{2}-\omega_{s}^{2}}\:x_{1}-\omega \, t)},\qquad\quad v_{5}^{+}=\beta_{5}\,e^{i(1/c_{m}\:\sqrt{\omega^{2}-\omega_{r}^{2}}\:x_{1}-\omega \, t)},\qquad\quad v_{6}^{+}=\beta_{6}\,e^{i(1/c_{m}\:\sqrt{\omega^{2}-\omega_{s}^{2}}\:x_{1}-\omega \, t)}.\nonumber 
\end{align}

In these formulas $\overline{\alpha}_{1},\,\overline{\alpha}_{2},\,\overline{\alpha}_{3}\in\R$ are the amplitudes of the incident (longitudinal and transverse) waves traveling in the Cauchy continuum that are assumed to be known. Moreover $\alpha_{1},\,\alpha_{2},\,\alpha_{3}\in\R$ are the amplitudes of the longitudinal and transverse waves which are reflected in the Cauchy medium once the incident wave reaches the interface. Analogously, $\beta_{1}^{1} ,\,\beta_{1}^{2}\in\R$ are the amplitudes associated to the longitudinal wave transmitted in the relaxed medium, while $h_{1}^{1},\,h_{1}^{2}\in\R^3$ and $k_{1}^{1},\,k_{1}^{2}\in\R$ are the eigenvectors and eigenvalues associated to the eigenvalue problem for longitudinal waves (see \cite{madeo2016reflection}). In the same way $\beta_{\alpha}^{1},\,\beta_{\alpha}^{2}\in\R,\,h_{\alpha}^{1},\,h_{\alpha}^{2}\in\R^3,\,k_{\alpha}^{1},\,k_{\alpha}^{2}\in\R\ (\alpha=2,3)$ are defined for the transverse waves transmitted in the relaxed medium (see \cite{madeo2016reflection} for details). Finally $ \beta_{4},\,\beta_{5},\,\beta_{6}\in\R$ are the amplitudes of the uncoupled waves transmitted in the relaxed medium.

Assuming the amplitudes of the incident waves to be known, we can count the 12 unknown amplitudes $\alpha_{1},\,\alpha_{2},\,\alpha_{3},\,\beta_{1}^{1} ,\,\beta_{1}^{2},\,\beta_{\alpha}^{1},\,\beta_{\alpha}^{2}\ (\alpha=2,3),\,\beta_{4},\,\beta_{5},\,\beta_{6}$ which can be calculated imposing the 12 scalar jump condition \eqref{Jump1}, \eqref{Jump2} and \eqref{Jump3}.

\subsection{Reflection and transmission of purely longitudinal waves at a Cauchy/relaxed-micromorphic interface}
In the remainder of this paper, we are interested in a first calibration of the constitutive parameters of our relaxed micromorphic model on a real experiment of wave transmission in a band gap metamaterial.

To do so, we focus on the experiment proposed in \cite{lucklum2012two} in which only longitudinal waves are considered. We hence consider here the solution of our relaxed problem only for what concerns the longitudinal part. In other words we are only considering the longitudinal fields $v_1$ and $u_1$ together with the field $v_{6}$ which is coupled to $v_1$ through the boundary conditions \eqref{Jump3}.

\begin{figure}[H]
	\begin{center}
		\begin{picture}(400,125)
		
		\put(18,89){\color{red}{incident longitudinal wave}}		
		\put(20,85){\color{red}{\vector(1,0){110}}}
		\put(28,72){\color{red}{$u_1^i=\overline{\alpha}_1 \,e^{i(\omega/c_{l}\,x_{1}-\omega \, t)} $}}
		
		\put(18,47){\color{red}{reflected longitudinal wave}}				
		\put(130,43){\color{red}{\vector(-1,0){110}}}
		\put(28,30){\color{red}{$u_1^r=\alpha_1 \,e^{i(-\omega/c_{l}\,x_{1}-\omega \, t)} $}}

		\put(200,89){\color{red}{transmitted longitudinal waves}}
		\put(170,85){\color{red}{\vector(1,0){210}}}
		\put(180,72){\color{red}{$v_1=\beta_{1}^{1} h_{1}^{1}e^{i(k_{1}^{1}(\omega)x_{1}-\omega \, t)}+\beta_{1}^{2} h_{1}^{2}e^{i(k_{1}^{2}(\omega)x_{1}-\omega \, t)}$}}

		\put(200,47){\color{Green}{transmitted uncoupled wave}}	
		\put(170,43){\color{Green}{\vector(1,0){210}}}
		\put(210,30){\color{Green}{$v_{6}^{+}=\beta_{6}\,e^{i(1/c_{m}\:\sqrt{\omega^{2}-\omega_{s}^{2}}\:x_{1}-\omega \, t)}$}}
		
		\put(133,0){$x_1=0$}
		\put(360,0){$\alpha=2,3$}
		\put(40,115){Cauchy medium $\ominus$}		
		\put(200,115){relaxed micromorphic medium $\oplus$}
		
		\put(0,15){\line(0,1){95}}
		\put(150,15){\line(0,1){95}}
		\put(400,15){\line(0,1){95}}
		\put(0,15){\line(1,0){400}}
		\put(0,110){\line(1,0){400}}
		\end{picture}
	\end{center}
	\caption{\label{FigWavesLong} Incident, reflected and transmitted longitudinal waves at a Cauchy/relaxed-micromorphic interface.}
\end{figure}
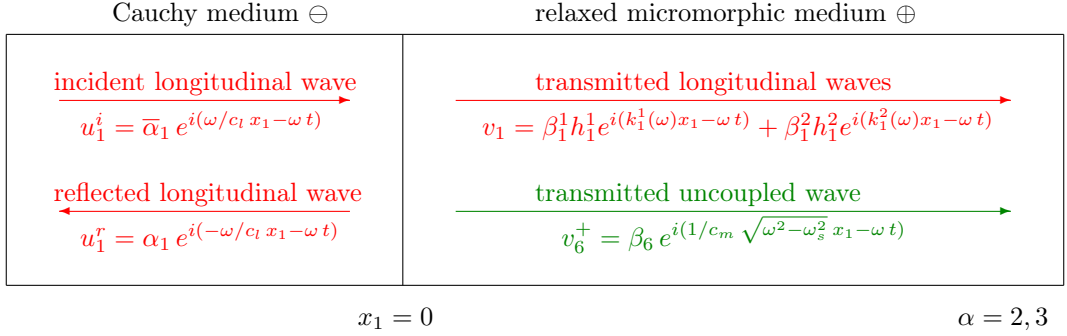

In summary, \textbf{the boundary value problem} for longitudinal waves can be written as (see equations  \eqref{CauchyAS}, \eqref{eq:BulkEq},  \eqref{Jump1}, \eqref{Jump2} and \eqref{Jump3})
\begin{align}
u_{1,tt}^{-}&=c_{l}^{2} (u_{1}^{-})'',
\nonumber\\
\nonumber\\
v_{1,tt}^{+}  &=A\,_{1}^{R}\cdot (v^{+}_1)'+ B\,_{1}^{R}\cdot (v_{1}^{+})'+ C\,_{1}^{R}\cdot v_{1}^{+},
\\\nonumber\\
\nonumber
v_{6,tt}^{+}&=A_{6}^{R} (v_{6}^{+})''+C_{6}^{R} v_{6}^{+},
\end{align}
together with the boundary conditions:
\begin{align}
 v_{1}^{+}\cdot\n-u_{1}^{-}&=0,
 \nonumber\\
 \nonumber\\
 \left(\begin{array}{c}
 \lle+2\me\\
 0\\ 0 \end{array}\right)\cdot\left( v_{1}^{+}\right)'&+\left(\begin{array}{c}
 0\\ -2\me\\ -(3\lle+2\me) \end{array}\right) \cdot v_{1}^{+}=(\lambda+2\mu)\,\left(u_{1}^{-}\right)'
\label{waveBC} \\\nonumber\\
\nonumber
 \left(\begin{array}{c}
 0\\ -\mLc/2\\ \mLc \end{array}\right)\cdot\left( v_{1}^{+}\right)'&=0, \qquad\qquad\qquad\quad\,
 \left(v_{6}^{+}\right)'=0.
\end{align}
The wave form solution for purely longitudinal fields is given by (see also equations \eqref{CauchyWaves}, \eqref{eq:WaveForm-1} and Figure \ref{FigWavesLong}):
\begin{align}
u_{1}^{-}&=\overline{\alpha}_{1}\,e^{i(\omega/c_{l}\,x_{1}-\omega \, t)}+\alpha_{1}\,e^{i(-\omega/c_{l}\,x_{1}-\omega \, t)},\vspace{1.2mm}
\nonumber\\
v_{1}^{+}&=\beta_{1}^{1} h_{1}^{1}e^{i(k_{1}^{1}(\omega)x_{1}-\omega \, t)}+\beta_{1}^{2} h_{1}^{2}e^{i(k_{1}^{2}(\omega)x_{1}-\omega \, t)},
\label{wavesol}\\ v_{6}^{+}&=\beta_{6}\,e^{i(1/c_{m}\:\sqrt{\omega^{2}-\omega_{s}^{2}}\:x_{1}-\omega \, t)}.\nonumber 
\end{align}

Replacing the wave solution \eqref{wavesol} in the 4 scalar jump conditions \eqref{waveBC} and setting $x_1=0$ (position of the interface) we can calculate the 4 unknown amplitudes $\alpha_1,\,\beta_{1}^{1} ,\,\beta_{1}^{2}$ and $\beta_6$.

From the condition  $\left(v_{6}^{+}\right)'=0$ it is straightforward to prove that $\beta_6=0$, so that finally $v_{6}^{+}=0\ \forall x_1$ and $ \forall t$. As for the other amplitudes, they have more complicated expressions that we do not explicitly show here since it does not add any fundamental information to the reasoning.


\section[Reflection and transmission coefficients at a Cauchy/relaxed-micromorphic interface]{Reflection and transmission coefficients at a Cauchy/relaxed-\\micromorphic interface}

We now want to define the reflection and transmission coefficients for the considered Cauchy/relaxed-micromorphic interface. To this purpose we introduce the quantities
\[
J_{i}=\int_{0}^{\Pi}H_{i}\left(0,t\right)dt,\qquad\qquad J_{r}=\int_{0}^{\Pi}H_{r}\left(0,t\right)dt,\qquad\qquad J_{t}=\int_{0}^{\Pi}H_{t}\left(0,t\right)dt,
\]
where $\Pi$ is the period of the traveling plane wave and $H_{i}\,$,
$H_{r}$ and $H_{t}$ are the energy fluxes of the incident, reflected
and transmitted energies, respectively. The reflection and transmission
coefficients can hence be defined as
\begin{equation}
R=\frac{J_{r}}{J_{i}},\qquad T=\frac{J_{t}}{J_{i}}.\label{eq:Reflection_coefficient}
\end{equation}
Since the considered system is conservative, one must have $R+T=1$.

In the particular case of reflection and transmission of longitudinal plane waves at a Cauchy/relaxed-micromorphic interface, recalling equations \eqref{Flux-Cauchy-1} and \eqref{eq:FluxRelaxed} together with the solutions \eqref{wavesol} for the unknown fields we have:
\begin{align}
H_{i}&=\dot{u}_{1}^{i} (\lambda+2\,\mu)\, u_{1,1}^{i} ,
\qquad\qquad
H_{r}=\dot{u}_{1}^{r} (\lambda+2\,\mu) \,u_{1,1}^{r} ,
\qquad\qquad
H_{t}=H_1^{1}+H_1^{6},
\end{align}
where we set $u_{1}^{i}=\overline{\alpha}_{1}\,e^{i(\omega/c_{l}\,x_{1}-\omega \, t)}$ and $u_1^r=\alpha_{1}\,e^{i(-\omega/c_{l}\,x_{1}-\omega \, t)}$. We explicitly remark that the fluxes $H_1^{1}$ and $H_1^{6}$ defined in equation \eqref{Fluxes} must be calculated with the solutions $v_1^{+}$ and $v_6^{+}$ obtained for the considered constraint and given in equation \eqref{wavesol}.

Since in this particular case we have shown that $v_{6}^{+}$ is zero, then $H_1^6$ does not contribute to the evaluation of the transmitted energy. Once the solution for the energy fluxes have  been calculated for the considered constraint (macro internal clamp with free microstructure), the reflection and transmission coefficients can be computed by using equations \eqref{eq:Reflection_coefficient}. We remark that R and T depend on the frequency $\omega$ of the traveling waves and that we must always have $R+T=1$.

	\subsection{Transmission coefficient at a Cauchy/relaxed-micromorphic interface: the degenerate limit case $L_c=0$ (internal variable model) \label{TransmInt}}						
We show here that at the interface between a Cauchy continuum and a relaxed micromorphic one it is possible to model, as a degenerate limit case, the onset of two band gaps whose bounds can be identified to be (see \cite{madeo2016reflection}): $[\omega_{l}^{1},\omega_{s}]$ and $[\omega_{l}^{2},\omega_{p}]$, where:
	\begin{align}
	\nonumber\omega_{l}^{1}&=\sqrt{\frac{a- \sqrt{a^{2}-b^{2} }}{2\eta\, (\lle+2\me)}},     \qquad \qquad \qquad \qquad\omega_{s}=\sqrt{\frac{2(\me+\mh)}{\eta}}     ,\\\label{Freq}	\\
	\nonumber\omega_{l}^{2}&=\sqrt{\frac{a+ \sqrt{a^{2}-b^{2}}}{2\eta\, (\lle+2\me)}} ,       \qquad \qquad\qquad\qquad\omega_{p}=\sqrt{\frac{2(\me+\mh)+3(\lle+\lh)}{\eta}}     ,
	\end{align}
	in which we have defined:
	\begin{align}
	a&=2\me(3 \lh + 2 \me + 4  \mh) + \lle (3 \lh + 6 \me + 4 \mh),\nonumber\\ 
	b^{2}&=8 (\lle + 
	2 \me) (\lle (3 \lh (\me + \mh) + 
	2 \mh (3 \me + \mh)) \\\nonumber &\qquad+ 
	2 \me (2 \mh (\me + \mh) + \lh (\me + 
	3 \mh))).
	\end{align}

\begin{figure}[H]
		\begin{centering}
			\begin{picture}(300,240)
			\put(0,0){\includegraphics[width=10cm]{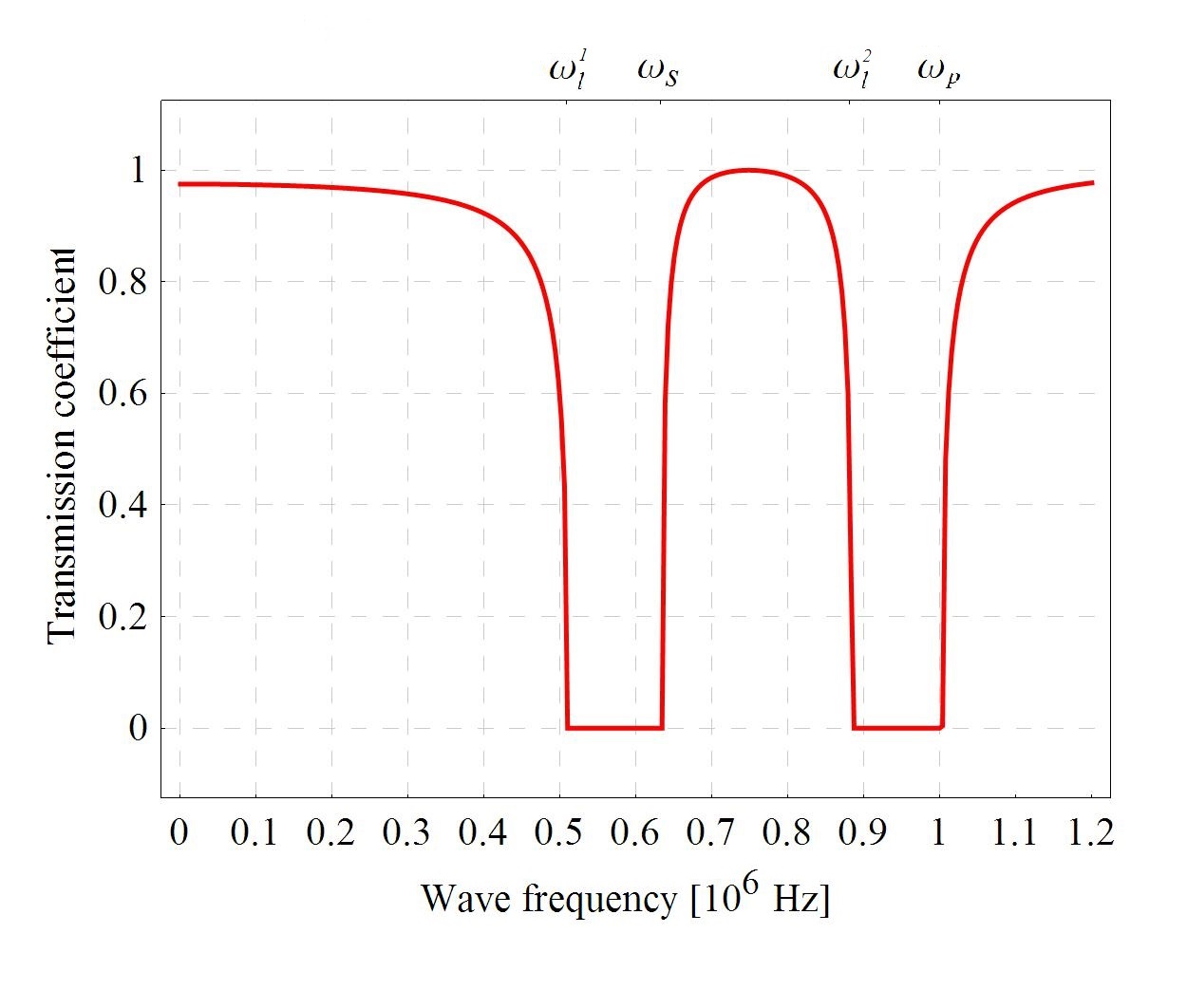}}

			\end{picture}

			\par\end{centering}

		\protect\caption{Transmission coefficient of the relaxed micromorphic model for   $ \lle=\me=\lh=\mh=100\ GPa$ and $\eta=1\ kg/m $.     \label{fig:TrasmissionRel}}
	\end{figure}

	In Figure \ref{fig:TrasmissionRel} we show a characteristic pattern of the transmission coefficient at a Cauchy/relaxed-micromorphic interface for a particular choice of the constitutive parameters and setting $L_c=0$. The main characteristic feature of the relaxed micromorphic model with $L_c=0$ (internal variable model) is that two separate band gaps can be determined and their bounds can be explicitly defined as functions of the constitutive parameters of the model according to equations \eqref{Freq}.
	
Switching on and slowly increasing the parameter $L_c$ produces small changes on the reflection profile of Figure \ref{fig:TrasmissionRel} basically related to the smoothening of the sharp corners that can be seen corresponding to the band-gap frequencies. We have shown in \cite{madeo2016reflection}	that such situation in which two band gaps are precisely identified with lower bounds  $\omega_{l}^{1}$ and   $\omega_{l}^{2}$, respectively, is a degenerate limit case of the relaxed micromorphic model obtained when setting the characteristic length $L_c$ to be identically zero. Such model with $L_c=0$ is also known as \textit{internal variable model} and it has been shown to be able to catch the main features of some particular classes of band-gap metamaterials (see \cite{sridhar2016homogenization,pham2013transient}). Nevertheless, the fact of completely ignoring non-local effects in materials with heterogeneous microstructure may induce a certain amount of inaccuracy in the modeling phase which could be hard to be controlled when necessary.

In the next section we will show that we can estimate the characteristic length $L_c$ of the metamaterial experimentally tested in \cite{lucklum2012two} to be comparable to the order of magnitude of the diameters of the embedded microstructure.

We will also show that, even if the estimated value of $L_c$ is non- negligible with respect to the characteristic size of the embedded microstructure, its effect on the amount of energy which is  transmitted in the considered metamaterial is very small. This means that the error which is introduced if one uses an internal variable model instead of a relaxed micromorphic one is energetically small. On the other hand, non-locality is a fundamental feature of metamaterials with heterogeneous microstructure and as such it should always be included in their modeling. As a matter of fact, non-local effects are sensible to become more and more important when the contrast in the mechanical properties between adjacent unitary cells at the microscopic level becomes more pronounced.

As a general rule, we claim that the degenerate limit case $L_c=0$ can be used for a first rough fitting  of the elastic parameters of the relaxed micromorphic model. After that, the characteristic length $L_c$ must be switched on in order to achieve a more accurate fitting of the experimental results. This last operation will allow for the estimate of non local effects in real metamaterials.

The relaxed micromorphic model allows for the possibility of including non-local effects in band-gap metamaterials. In the next section we will have the twofold task of:
	\begin{itemize}
		\item fitting at best our constitutive parameters on a real metamaterial,
		\item estimate the order of magnitude of non-localities in such metamaterial.
	\end{itemize}

\section{Modeling a two dimensional phononic crystal via the relaxed micromorphic model}
In this section we are interested in the modeling of the mechanical behavior of a particular metamaterial (phononic crystal) which has been seen to inhibit elastic wave propagation on an experimental basis (see \cite{lucklum2012two}).

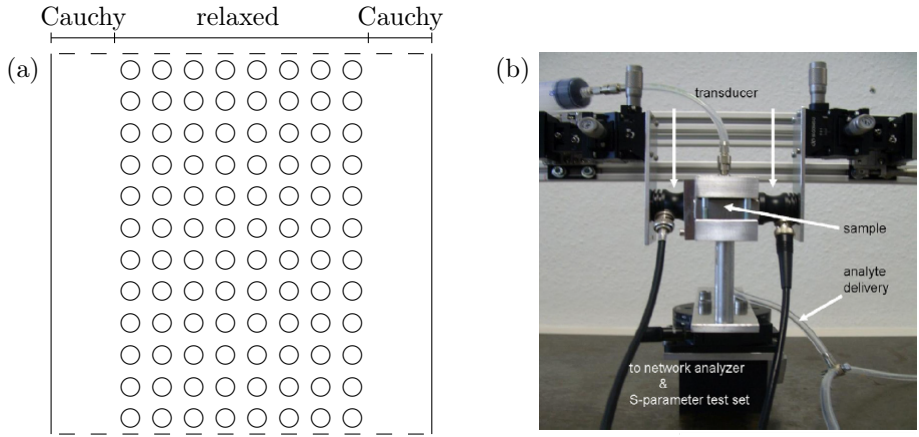
\begin{figure}[H]
	\begin{centering}

		\par\end{centering}
	\begin{picture}(180,170)
	\put(-17,135){(a)}	\put(0,0){\line(0,1){144}}
	\put(144,0){\line(0,1){144}}
	\multiput(3,0)(12,0){12}{\line(1,0){6}}
	\multiput(3,144)(12,0){12}{\line(1,0){6}}
	\multiput(30,6)(12,0){8}{
		\multiput(0,0)(0,12){12}{\circle{7.2}}}
	\put(168,135){(b)}
	
	\put(-4,155){Cauchy}		
	\put(55,155){relaxed}
	\put(116,155){Cauchy}	

	\put(0,150){\line(1,0){144}}
	\multiput(0,148)(144,0){2}{\line(0,1){4}}
	\multiput(24,148)(96,0){2}{\line(0,1){4}}
	\end{picture}
		\includegraphics[height=5.12cm]{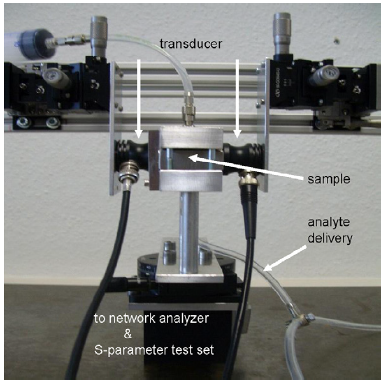}

		\protect\caption{Schematics of the sample structure (a) and the experimental setup (b) (\cite[Figure 1(b)]{lucklum2012two}\label{fig:Experimental}).}
		\end{figure}

The structure presented, which is schematically shown  in Figure \ref{fig:Experimental}(a) consists of a steel plate with liquid-filled holes in square array. The lattice constant, denoted as a, is 3.0 mm, the thickness t of the plate is 15 mm, the diameter d of the hole is 1.8 mm and the width of the cavity, w, is 1.5 mm.

\subsection{Experiments of wave transmission at a Cauchy/phononic-crystal interface}
\begin{figure}[H]
			\begin{centering}
				\includegraphics[width=10cm]{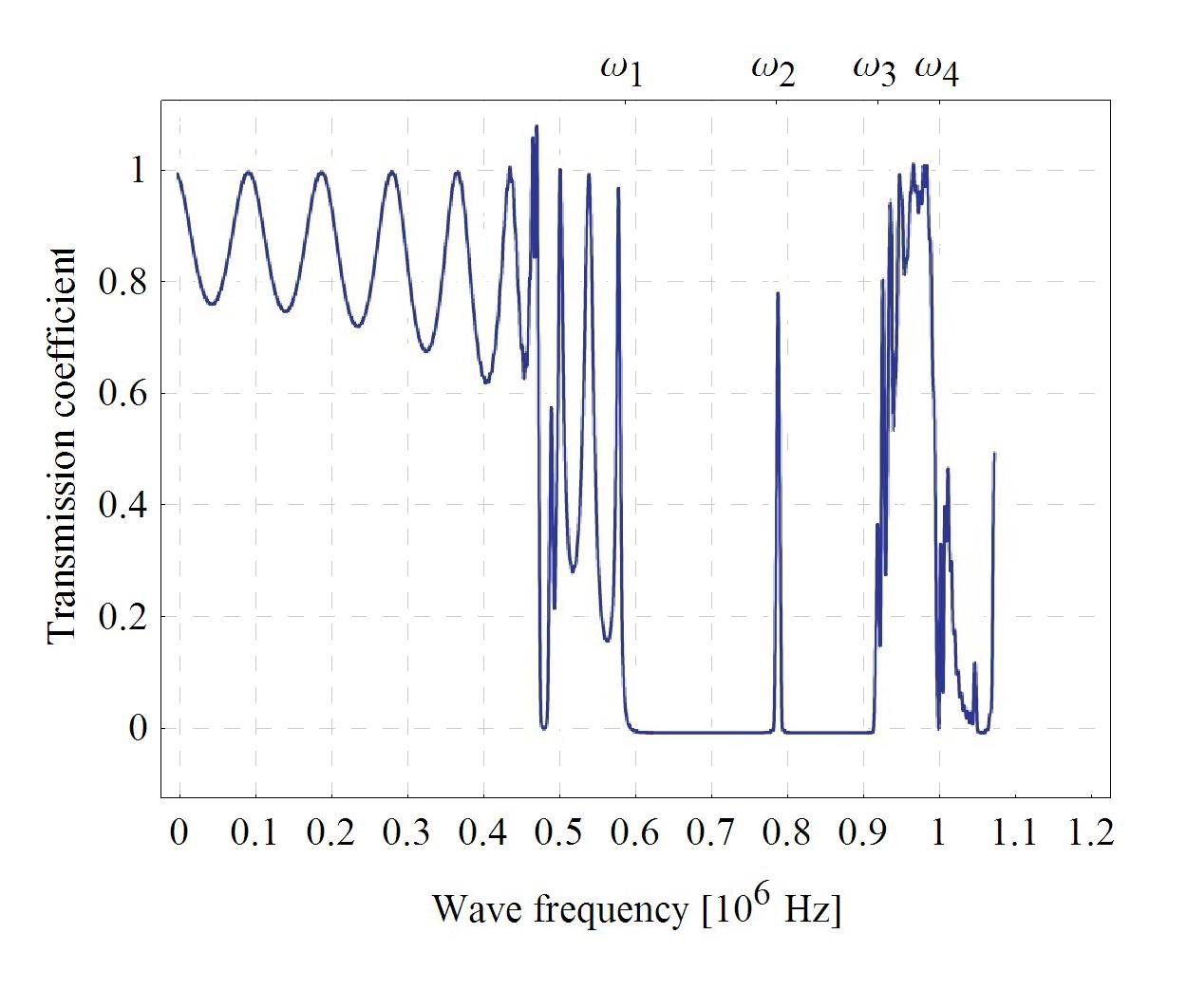}
				\par\end{centering}
				
				\protect\caption{Transmission spectrum of the phononic crystal presented in Figure \ref{fig:Experimental}a, with liquid-filled
					holes \cite[Figure 2b]{lucklum2012two}.\label{fig:Trasmission}}
					\end{figure}

		We show in Figure \ref{fig:Trasmission}  the obtained experimental transmission spectrum of the considered phononic crystal, i.e. with 8 rows of liquid-filled holes (see Figure \ref{fig:Experimental}(a)) as a function of the frequency of the traveling wave. Given the geometry of the specimen shown in Figure \ref{fig:Experimental}, a longitudinal wave is sent in the Cauchy medium left side and the transmission coefficient is evaluated when the wave leaves the metamaterial on the opposite side.
		 With liquid filled holes the band gap edge crosses the $-3$ dB-level at $\omega_{1}=586 kHz$. Transmission of acoustic waves is suppressed until the upper edge at $\omega_{3}=918$ kHz but a single peak arises $\omega_{2}=793$ kHz, which can be attributed to the resonance of the liquid-filled holes. The periodic variation of transmission at lower frequencies is caused by Bragg resonances. The second transmission band extends to about $\omega_{4}\simeq1$ MHz. 
		
	\subsection{Identification of the parameters}		
	In this subsection we present the procedure that we used in order to fit in the best possible way the maximum possible number of parameters of our relaxed micromorphic model on the available data based on a real phononic crystal. To start with, we fix the macroscopic mass density to be known as the averaged density of steel with fluid-filled holes. In particular we choose $\rho=5000\ kg/m^3$. Nevertheless, we verify a posteriori that the value of $\rho$ indeed does not sensibly affect the profile of the reflection coefficient for frequencies between 0 and $1\ MHz$. This fact is sensible if, with reference to \cite{madeo2015wave,madeo2016reflection} and to Figure \ref{CurlNon}, we notice that the parameter $\rho$ only intervenes in the definition of the oblique asymptote $c_p=\sqrt{\frac{\lle+2\me}{\rho}}$ for longitudinal waves. Such asymptote governs the slope of the optic wave LO1 which starts playing a significant role for frequencies higher than $\omega_p$. In the considered example $\omega_p$ will be set to be equal to $\omega_3$ which is experimentally seen to be close to $1\ MHz$. For frequencies higher than $1\ MHz$ variations of $\rho$ could eventually produce more tangible changes in the profile of the reflection coefficient. 
	
	To perform the fitting of the remaining parameters, we started by imposing the following identities:
	\begin{align}
\omega_{l}^{1}(\me,\mh,\lle,\lh,\eta)&=\omega_{1}, &\omega_{p}(\me,\mh,\lle,\lh,\eta)&=\omega_{3},\label{system}\\
\omega_{l}^{2}(\me,\mh,\lle,\lh,\eta)&=\omega_{2},
&\omega_{s}(\me,\mh,\eta)&=\omega_{2}-8\,kHz.\nonumber
	\end{align}
where we recall that the explicit expression of $\omega_{l}^{1},\,\omega_{s},\,\omega_{l}^{2},\,\omega_{p}$ as function of the elastic parameters of the relaxed micromorphic model is given in equations \eqref{Freq}. We hence have 4 conditions to determine the 5 elastic parameters $\me,\mh,\lle,\lh,\eta$ setting in a first instance $L_c=0$. If analogous experiments as the one proposed in \cite{lucklum2012two} for longitudinal waves would be reproduced on the same metamaterial but for transverse waves, extra conditions on the parameters of the relaxed micromorphic model would be available that would permit a more accurate fitting.

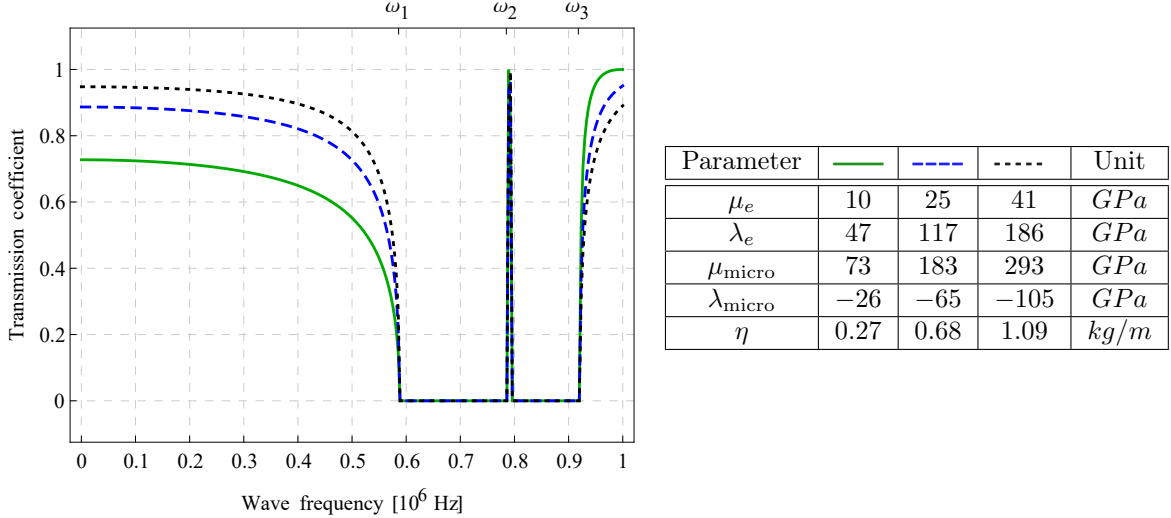
\begin{figure}[H]
		\begin{centering}
				\begin{picture}(450,220)
				\put(0,20){
			\includegraphics[height=7cm]{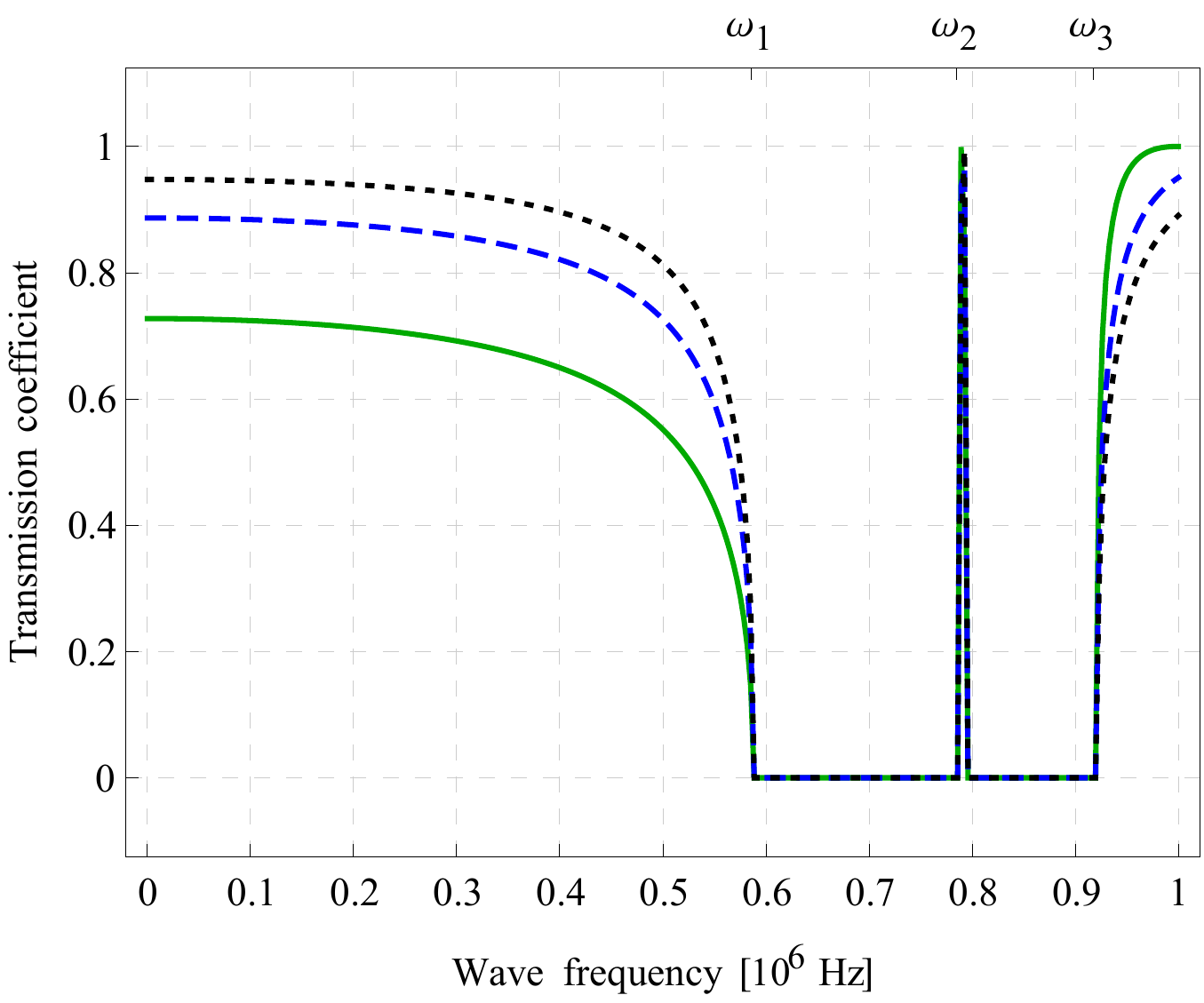}}	
		
		\put(250,120){
				\begin{tabular}{|c|c|c|c|c|}
					\hline 
					Parameter & &  & & Unit\tabularnewline
					\hline 		\hline 
					$\me$ & $10$& $25$& $41$ & $GPa$\tabularnewline
					\hline 
					$\lle$ & $47$& $117$ & $186$ & $GPa$\tabularnewline
					\hline 
					$\mh$ & $73$& $183$& $293$ & $GPa$\tabularnewline
			
					\hline 
					$\lh$ & $-26$& $-65$& $-105$ & $GPa$\tabularnewline
					\hline 
					$\eta$ & $0.27$& $0.68$& $1.09$ & $kg/m$\tabularnewline
					\hline 
				\end{tabular}}
				\multiput(317,154)(0,0.3){3}{
				\put(0,0){\color{Green}{\line(1,0){19}}}
				\multiput(30,0)(5,0){4}{\color{blue}{\line(1,0){4}}}
				\multiput(61,0)(4,0){5}{\color{black}{\line(1,0){2}}}}
			\end{picture}
			\par\end{centering}
			\caption{\label{fig:Par}Profiles of the transmission coefficients obtained for different arbitrary values of the parameter $\me$. The other parameters corresponding to any given value of $\me$ are obtained according to the solution \eqref{sol} and are shown in the table given on the right. }
	\end{figure}

We start by numerically solving the system of four equations \eqref{system} with respect to the parameters $\lle, \mh,\lh$ and $\eta$ leaving free the parameter $\me$. The obtained solution is\footnote{We explicitly mention that, additionally to the solution \eqref{sol} we obtain a second solution which, nevertheless must be excluded since it violates the positive definiteness of the strain energy density $W$. Solution \eqref{sol}  is then the only possible solution which can be used to fit the profile of the transmission coefficient. We checked that it is possible to leave free any other parameter rather than $\me$ to perform the desired fitting of the transmission coefficient and that it yields comparable results for the obtained values of the constitutive parameters.}
	\begin{align}
	\lle&=4.58 \,\me, \qquad
	\mh=7.21  \,\me, \qquad
	\lh=-2.57 \, \me, \qquad
	\eta=2.66\ 10^{-11} \,\me, \label{sol}
	\end{align}
The free parameter $\me$ is then varied in order to evaluate its influence on the reflection coefficient. A parametric study on the free coefficient $\me$ is performed giving rise to the profiles of the transmission coefficients shown in Figure \ref{fig:Par}.

At this point, we are able to choose the value of the parameter $\me$ which respects the conditions \eqref{sol} and which fits at best the profile of Figure \ref{fig:Trasmission}. We conclude that, based on the described fitting procedure, the values of the parameters that best fits the profile associated to the real phononic crystal are those presented in Table \ref{ParametersValuesInt}b.

\begin{table}[H]
	\begin{centering}
\begin{tabular}{c}(a)\tabularnewline\tabularnewline\tabularnewline\tabularnewline \tabularnewline \tabularnewline 	\end{tabular}	\begin{tabular}{|c|c|c|}
			\hline 
			Condition & Value & Unit\tabularnewline
			\hline 
			\hline
			\hline
			$\omega_{l}^{1}=\omega_{1}$ & $586$ & $kHz$\tabularnewline
			\hline 
			$\omega_{s}=\omega_{2}-8\,kHz$ & $785$ & $kHz$\tabularnewline
			\hline 
			$\omega_{l}^{2}=\omega_{2}$ & $793$ & $kHz$\tabularnewline
			\hline 
			$\omega_{p}=\omega_{3}$ & $918$ & $kHz$\tabularnewline
			\hline 
			
		\end{tabular}\quad{}\quad{}\quad{}\quad{}%
\begin{tabular}{c}(b)\tabularnewline\tabularnewline\tabularnewline\tabularnewline\tabularnewline\tabularnewline 	\end{tabular}	
		\begin{tabular}{|c|c|c|}
			\hline 
			Parameter & Value & Unit\tabularnewline
			\hline 	\hline 
			$\me$ & $25$ & $GPa$\tabularnewline
			\hline 
	$\lle$ & $117$ & $GPa$\tabularnewline
		
		\hline 
		$\mh$ & $183$ & $GPa$\tabularnewline
			\hline 
	$\lh$ & $-65$ & $GPa$\tabularnewline
			\hline 
	$\eta$ & $0.68$ & $kg/m$\tabularnewline
			\hline 
		\end{tabular}
		\par\end{centering}
	\caption{\label{ParametersValuesInt}Conditions used for the parameters identification (a) and corresponding values of the obtained elastic parameters of the relaxed micromorphic model (b).}
\end{table}	

\begin{figure}[H]
		\begin{centering}

				\includegraphics[height=9cm]{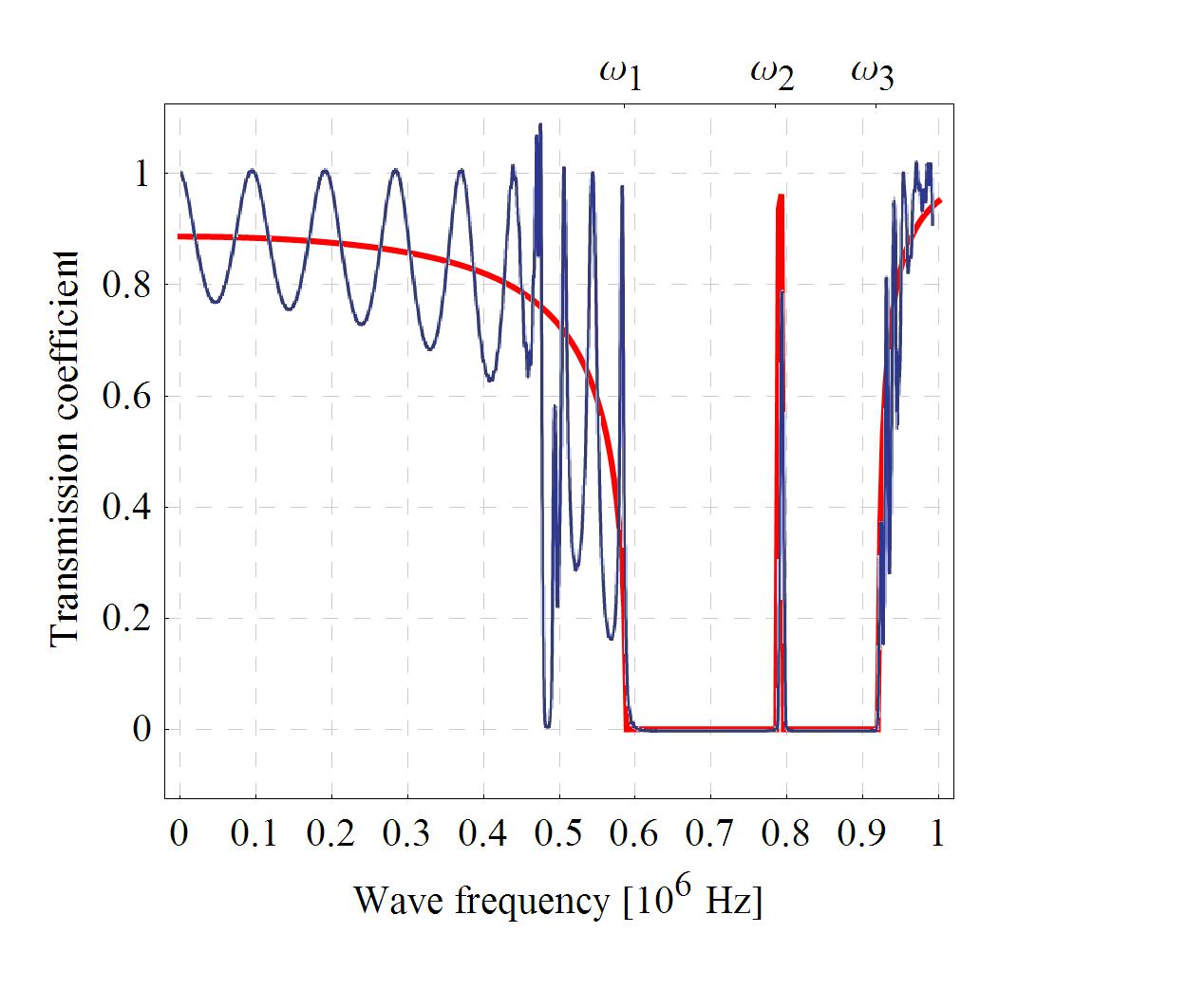}	
			\par\end{centering}
		\caption{\label{fig:Chosen} Comparison of the profile obtained in \cite{lucklum2012two} based on a real metamaterial and the one obtained with the relaxed micromorphic model with the values given in Table \ref{ParametersValuesInt} and $L_c=0$. }
	\end{figure}

The corresponding profile of the transmission coefficients as compared to the one presented in \cite{lucklum2012two} is given in Figure \ref{fig:Chosen}. 

Figure \ref{fig:Chosen} shows the comparison between the profile of the transmission coefficient obtained in \cite{lucklum2012two} for a real phononic crystal and the one obtained with our relaxed micromorphic model when setting $L_c=0$. 

It can be seen that a very good fitting can be obtained up to frequencies of the order of $1\ MHz$. 

In particular, the oscillatory behavior observed for lower frequencies and which, according to the authors of \cite{lucklum2012two}, is due to Bragg scattering phenomena is catched by our model in an ``averaged'' sense. 

The fitting for higher frequencies is almost perfect up to reaching $1\ MHz$, while for frequencies higher than $1\ MHz$ the relaxed micromorphic model looses it predictivity due to the fact that the corresponding wavelengths are so small that the continuum hypothesis is sensible to become inaccurate.

We need to explicitly remark that the peak of reflection, which is obtained around the frequency $\omega_2$ and that is experimentally related to a resonant behavior of the fluid inside the walls is slightly overestimated by the simulation via the relaxed micromorphic model with respect to the one observed in \cite{lucklum2012two}. This peak magnification can be related to the fact that no dissipation is accounted for in our model, while the fluid viscosity may perhaps play here a non-negligible role. 

We now come back to the point where we set $L_c=0$ in order to start fitting our constitutive parameters (see subsection \ref{TransmInt}). This fact allowed us to obtain here the values of the elastic parameters of our model by a first fitting with the profile of the transmission coefficient (see Table \ref{ParametersValuesInt}).

\begin{figure}[H]
		\begin{centering}
			\begin{picture}(450,200)
			\put(0,-18){\includegraphics[height=7cm]{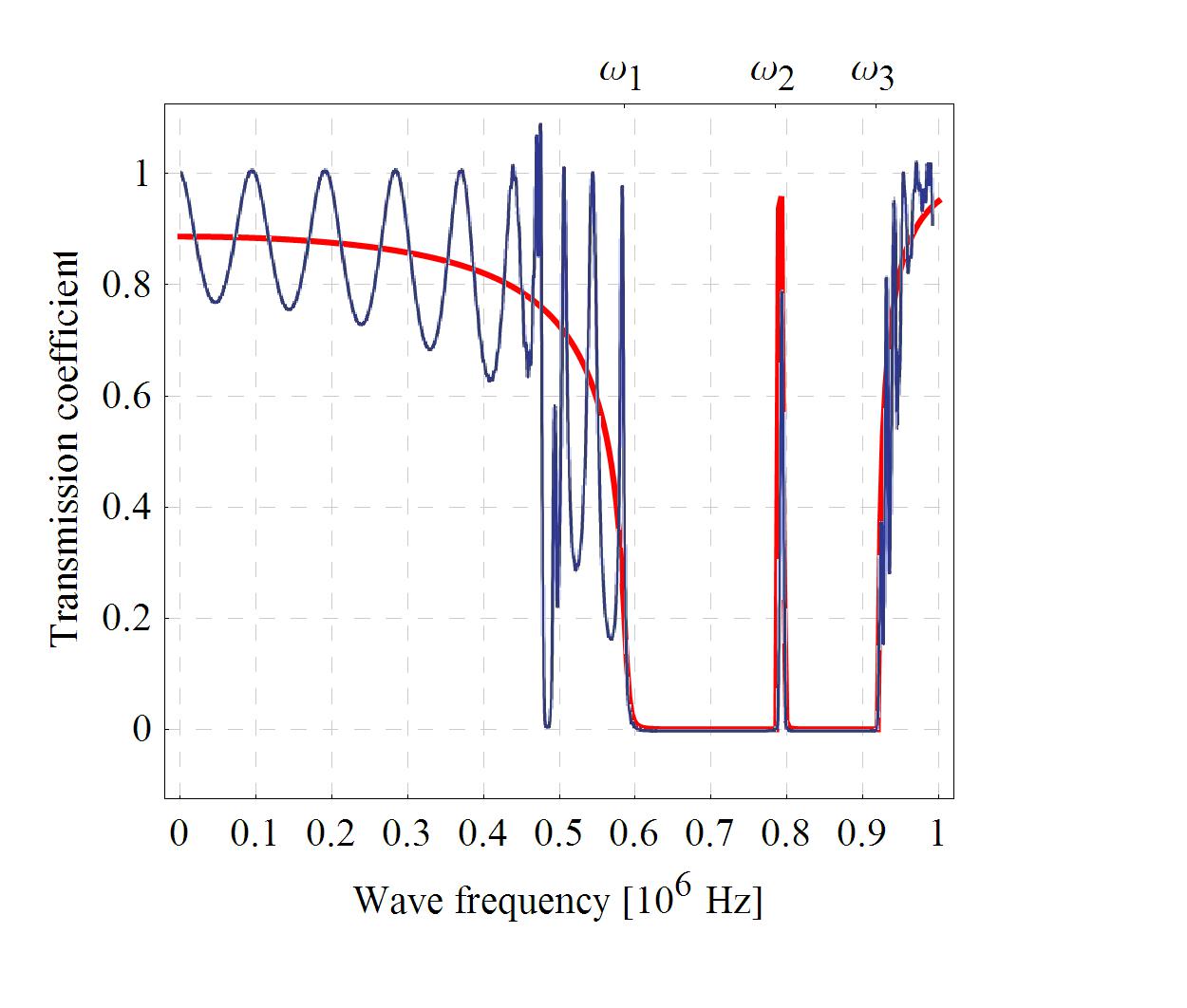}}
			
			\put(220,0){\includegraphics[height=6cm]{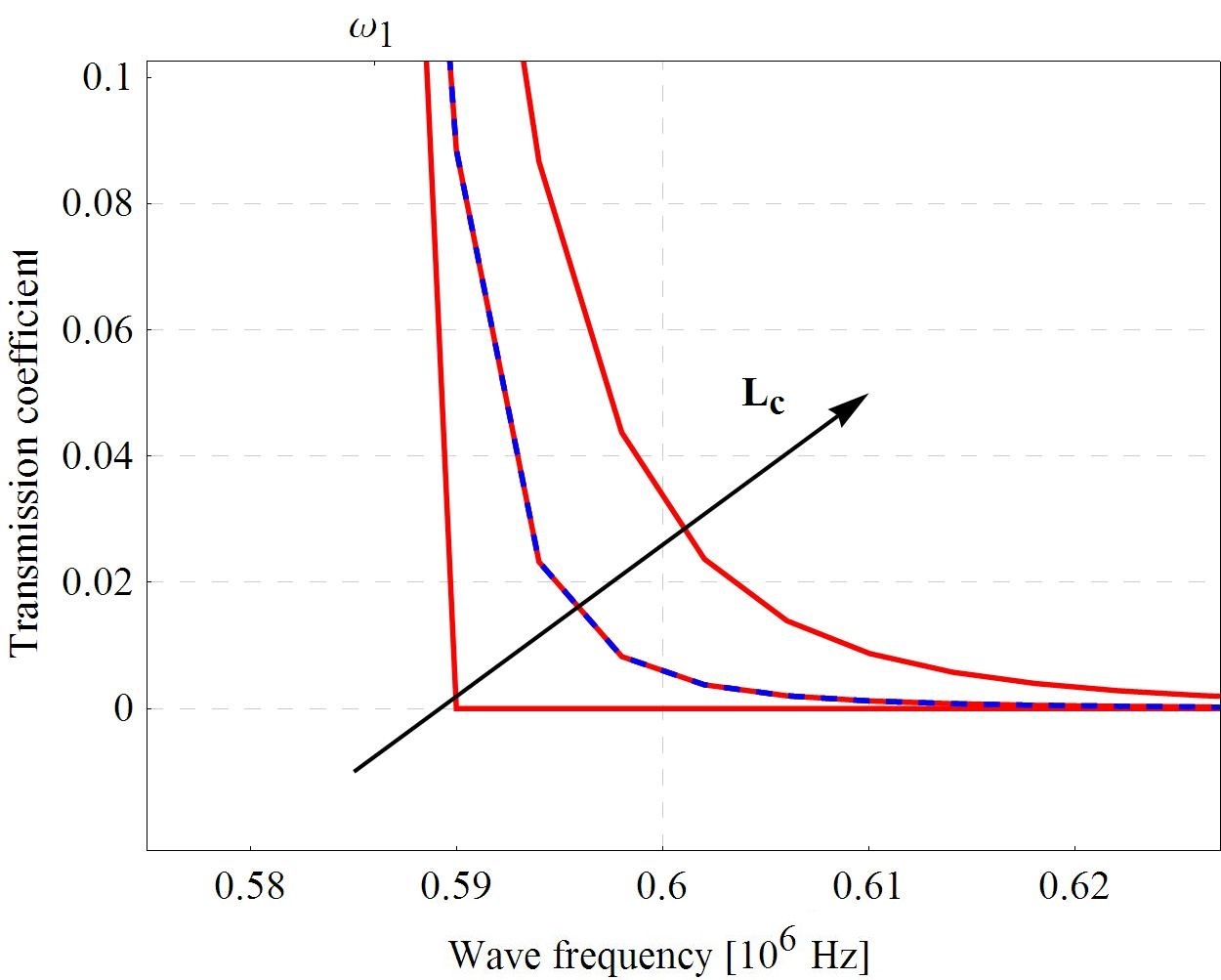}}
			\put(118,33){\multiput(0,0)(0,15){2}{\line(1,0){15}}
								\multiput(0,0)(15,0){2}{\line(0,1){15}}
								\put(16,0){\vector(1,0){110}}}
			\end{picture}

			\par\end{centering}
		
		\protect\caption{Fitting of the parameter $L_c$ on the experimental profile of the transmission coefficients.\label{fig:TrasmissionSim}}
	\end{figure}
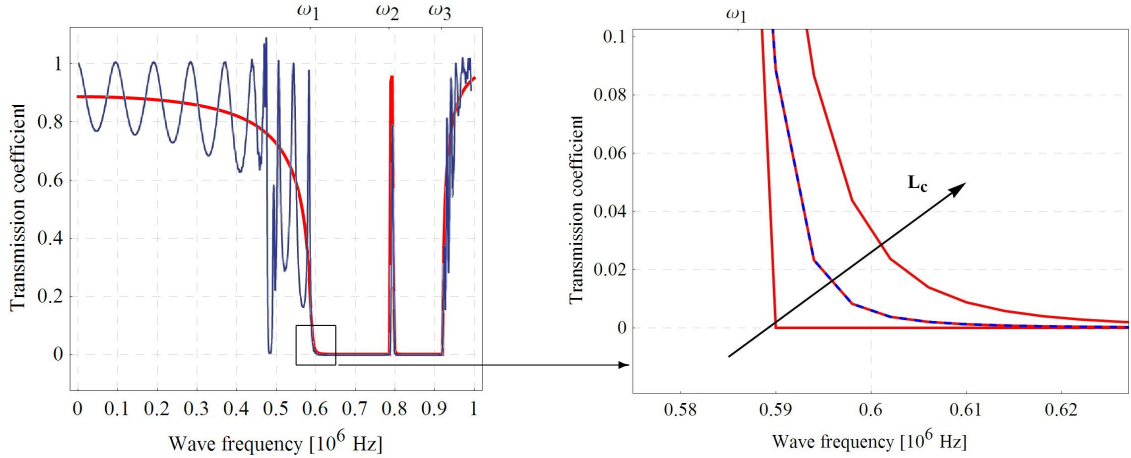

On the other hand, as expected, switching on the characteristic length $L_c$ allows an even better fitting as shown in Figure \ref{fig:TrasmissionSim}.

Indeed, we can notice from Figure \ref{fig:TrasmissionSim} that the degenerate limit case $L_c=0$ lets the calculated transmission coefficient slightly deviate from the experimental one (sharp corners). Small variations of the numerical profile can be perceived as far as $L_c \in [0,0.5\, mm)$. On the other hand, as far as $L_c = 0.5\, mm$ an almost perfect fitting is achieved (dashed line in Figure \ref{fig:TrasmissionSim} on the right)). This means that we have been able to estimate the non-locality of the considered metamaterial to be of the order of $0.5\, mm$, i.e. $\sim1/3$ of the diameters of the holes.

We need to explicitly say that the value of the macroscopic density $\rho$ might slightly affect the variation of the transmission coefficient as function of $L_c$. Nevertheless, we need to consider a density of 1 order of magnitude higher ($50\,000\  kg/m^3$) in order to appreciate a sensible deviation of the profiles shown in Figure \ref{fig:TrasmissionSim}. We leave to a subsequent work the aim of determining also the macroscopic mass density $\rho$ by using extra conditions provided by the fact of considering also measurements on transverse waves.

The determination of the parameter $L_c$ completes the fitting of the elastic parameters of our relaxed micromorphic model on the band-gap metamaterial experimentally tested in \cite{lucklum2012two} (see also Table \ref{ParametersValuesInt}). The Cosserat couple modulus parameter $\mc$ cannot be measured as far as only longitudinal waves are considered here and it thus remains to be determined. We have to explicitly remark that if an analogous fitting procedure would have been possible for transverse waves, we would have had more conditions than parameters to be determined. The extra conditions could have been used as validation of the fitted parameters.

It is out of the scope of this paper to envisage a general procedure for the optimal fitting of the whole set of parameters of the relaxed micromorphic model. We leave this fundamental objective to a forthcoming paper where such general procedure will be deeply discussed based on experiments of real interest.

The main scope of the present paper that we think to have successfully achieved is threefold:
\begin{itemize}
	\item we give the very first estimation of the maximum possible number of constitutive parameters of the relaxed micromorphic model based on a simple measurement of transmission of longitudinal waves at a Cauchy/band-gap-metamaterial interface
	\item we give the very first evidence of the non-locality in band-gap metamaterials based upon real experiments
	\item we elucidate the physical meaning of the constraint which has been introduced in \cite{madeo2016reflection} and that we called ``\textit{internal clamp with free microstructure}'': such constraint allows for the description of continuity of displacement in the solid phase at the Cauchy/metamaterial interface, while the fluid in the embedded microstructure is free to vibrate. It is exactly the freedom which is left to the micro-motions that allows for the description of the local resonant peak around the frequency $\omega_2$ which is indeed not possible for other types of constraints (see \cite{madeo2016reflection}).
\end{itemize}

\section {Conclusions}
In the present paper we give the very first estimate of the elastic coefficients of the relaxed micromorphic model based upon the experimentally-based results presented in \cite{lucklum2012two} which concern the measurement of the transmitted energy as a function of the frequency of the traveling wave for a particular band-gap metamaterial. More particularly, restricting our attention to the problem of studying reflection and transmission of longitudinal waves at a Cauchy/relaxed-micromorphic interface, we are able to reproduce the main characteristic features which are observed in \cite{lucklum2012two} for a phononic crystal obtained by means of an aluminum plate with small fluid-filled holes (diameter $\sim 1.8\,mm$).

\medskip
Suitably choosing the values of the parameters of our relaxed micromorphic model, we are able to fit the profile of the transmission coefficient proposed in \cite{lucklum2012two} as function of the frequency of the traveling waves.

Two band-gaps which almost collapse to form a unique band-gap can be observed both in \cite{lucklum2012two} and as a result of the simulations based  upon our relaxed micromorphic model.

The continuity of such extended band-gap is broken due to the presence of a resonant peak of transmitted energy that is seen to be related to the internal resonance of the fluid embedded in the microstructure.

We present a detailed procedure that we use to fit almost all the parameters of our relaxed-micromorphic model except the Cosserat couple modulus $\mc$ which remains undetermined. This indeterminacy is due to the fact that experiments concerning reflection and transmission of transverse waves in the considered metamaterial have not been performed yet.

We leave to a forthcoming contribution the problem of determining the whole set of parameters of the relaxed micromorphic model for real band-gap metamaterials.

\medskip
The results presented in this  paper allow to give the first physical interpretation of the boundary conditions that can be imposed at a Cauchy/relaxed-micromorphic interface based upon a real experiment.

\medskip

We conclude our paper with the finding that we believe to be the most important to be pointed out. Indeed, we showed by direct comparison of our relaxed micromorphic model with available evidences that non-local effects are an intrinsic feature of band-gap metamaterials.

A characteristic length $L_c=0.5\, mm$ has been estimated for the real phononic crystal studied in \cite{lucklum2012two} which is almost 1/3 of the diameter of the holes in the embedded microstructure. 

Even if the energetic contribution associated to the underlying non-locality is very small (only small changes in the transmission coefficient can be appreciated when increasing $L_c$ from 0 to $0.5\,mm$), such non-locality is intrinsically present in any microstructured material and as such it should be always accounted for when modeling their mechanical behavior.

The macroscopic effects of non-localities are sensible to become more and more energetically significant when considering stronger contrasts in the mechanical properties at the microscopic level (e.g. unitary cells with very different stiffnesses). The relaxed micromorphic model should be always used when one wants to model band-gap metamaterials in order to account for such non-localities.

\medskip

In a subsequent work we will provide a more complete determination of the constitutive parameters of the relaxed micromorphic model for real band-gap metamaterials and we will discuss further the importance of non-local effects in such microstructured materials.

\section{Acknowledgments}

Angela Madeo thanks INSA-Lyon for the funding of the BQR 2016 \textquotedbl{}Caractérisation
mécanique inverse des métamatériaux: modélisation, identification
expérimentale des paramètres et évolutions possibles\textquotedbl{}.

\footnotesize

\let\stdsection\section
\def\section*#1{\stdsection{#1}}

\bibliography{library}
\bibliographystyle{plain}

\let\section\stdsection

\end{document}